\documentclass[12pt]{article}
\pdfoutput=1
\usepackage[top=1in, bottom=1in, left=1in, right=1in]{geometry}
\usepackage{latexsym}
\usepackage{amssymb, amsfonts, amsmath}
\usepackage{indentfirst}
\usepackage{bbm}
\usepackage{amssymb}
\usepackage{verbatim}
\usepackage{amsmath, amsthm, amssymb}
\usepackage{mathrsfs}
\usepackage{amsfonts}
\usepackage{dsfont}
\usepackage{csquotes}
\usepackage{caption}
\usepackage{hyperref}
\usepackage{url}
\usepackage{xcolor}
\usepackage{xpatch}
\usepackage{multirow}
\usepackage{diagbox}
\usepackage{booktabs}
\usepackage[english]{babel}

\usepackage{graphicx} 
\usepackage[export]{adjustbox}
\usepackage{float}
\graphicspath{ {images/} }

\def\smath#1{\text{\scalebox{.8}{$#1$}}}

\def\sfrac#1#2{\smath{\frac{#1}{#2}}}

\usepackage[style=phys, articletitle=true, biblabel=brackets, backend=bibtex,
chaptertitle=false, pageranges=false, eprint=true,%
natbib=true, maxbibnames=10, giveninits=true, sorting=none]{biblatex} 

\DeclareFieldFormat[article]{journaltitle}{\mkbibitalic{#1\isdot}} 
\makeatletter
\newrobustcmd{\mkbibfixedbrackets}[1]{%
	\begingroup
	\blx@blxinit
	\blx@setsfcodes
	\bibleftbracket#1\bibrightbracket
	\endgroup}

\renewbibmacro*{doi+eprint+url}{%
	\iftoggle{bbx:doi}
	{\printfield{doi}}
	{}%
	\ifboolexpr{test {\iffieldequalstr{eprinttype}{arxiv}} or test {\iffieldequalstr{eprinttype}{arXiv}}}
	{\setunit{\addspace}\newblock}
	{\newunit\newblock}%
	\iftoggle{bbx:eprint}
	{\usebibmacro{eprint}}
	{}%
	\newunit\newblock
	\iftoggle{bbx:url}
	{\usebibmacro{url+urldate}}
	{}}

\xpatchbibdriver{online}
{\newunit\newblock
	\iftoggle{bbx:eprint}
	{\usebibmacro{eprint}}
	{}}
{\ifboolexpr{test {\iffieldequalstr{eprinttype}{arxiv}} or test {\iffieldequalstr{eprinttype}{arXiv}}}
	{\setunit{\addspace}\newblock}
	{\newunit\newblock}%
	\iftoggle{bbx:eprint}
	{\usebibmacro{eprint}}
	{}}
{}{}

\DeclareFieldFormat{eprint:arxiv}{%
	\mkbibbrackets{%
		\ifhyperref
		{\href{http://arxiv.org/\abx@arxivpath/#1}{%
				arXiv\addcolon
				\nolinkurl{#1}%
				\iffieldundef{eprintclass}
				{}
				{\addspace\UrlFont{\mkbibfixedbrackets{\thefield{eprintclass}}}}}}
		{arXiv\addcolon
			\nolinkurl{#1}%
			\iffieldundef{eprintclass}
			{}
			{\addspace\UrlFont{\mkbibfixedbrackets{\thefield{eprintclass}}}}}}}
\makeatother

\parskip=\medskipamount

\DeclareMathAlphabet{\mathbbmsl}{U}{bbm}{m}{sl}
\DeclareMathAlphabet\mathbfcal{OMS}{cmsy}{b}{n}

\newcommand{\cA}{{\cal A}}

\newcommand{\cF}{{\cal F}}

\newcommand{\cH}{{\cal H}}
\newcommand{\cI}{{\cal I}}

\newcommand{\cN}{{\cal N}}

\newcommand{\cT}{{\cal T}}

\def\a{\alpha}

\def\b{\beta}

\def\d{\delta}
\def\e{\epsilon}
\def\f{\phi}
\def\g{\gamma}
\def\G{\Gamma}

\def\l{\lambda}

\def\s{\sigma}

\def\D{\Delta}
\def\F{\Phi}
\def\J{\Psi}

\def\P{\Pi}

\newcommand{\ad}{{\dot{\alpha}}}                           
\newcommand{\bd}{{\dot{\beta}}}                            
\newcommand{\ve}{\varepsilon}                            

\newcommand{\pa}{\partial}                           

\newcommand{\be}{\begin{equation}}
\newcommand{\ee}{\end{equation}}
\newcommand{\bea}{\begin{eqnarray}}
\newcommand{\eea}{\end{eqnarray}}

\newcommand{\ba}{\begin{array}}
\newcommand{\ea}{\end{array}}

\def\double #1{#1{\hbox{\kern-2pt $#1$}}}

\newcommand{\gd}{{\dot\g}}

\newcommand{\bsubeq}{\begin{subequations}}
\newcommand{\esubeq}{\end{subequations}}

\numberwithin{equation}{section}

\begin{document}

\begin{titlepage}
\begin{flushright}
July, 2023
\end{flushright}
\vspace{2mm}

\begin{center}
\Large \bf Three-point functions of conserved currents in 4D CFT: general formalism for arbitrary spins
\end{center}

\begin{center}
{\bf
Evgeny I. Buchbinder and Benjamin J. Stone}

{\footnotesize{
{\it Department of Physics M013, The University of Western Australia\\
35 Stirling Highway, Crawley W.A. 6009, Australia}} ~\\
}
\end{center}
\begin{center}
\texttt{Email: evgeny.buchbinder@uwa.edu.au,\\ benjamin.stone@research.uwa.edu.au}
\end{center}

\vspace{4mm}
\begin{abstract}
\baselineskip=14pt
We analyse the general structure of the three-point functions involving conserved higher-spin currents $J_{s} := J_{\a(i) \ad(j)}$ belonging to any Lorentz representation in 
four-dimensional conformal field theory. Using the constraints of conformal symmetry and conservation equations, we computationally analyse the general 
structure of three-point functions $\langle J^{}_{s_{1}} J'_{s_{2}} J''_{s_{3}} \rangle$ for arbitrary spins and propose a classification of the results. For bosonic vector-like currents with $i=j$, it is known that the number of independent conserved structures is $2 \min (s_{i}) + 1$. For the three-point functions of conserved currents with arbitrarily many dotted and undotted indices, we show that in many cases the number of structures deviates from $2 \min (s_{i}) + 1$.




\end{abstract}
\end{titlepage}

\newpage
\renewcommand{\thefootnote}{\arabic{footnote}}
\setcounter{footnote}{0}

\tableofcontents
\vspace{3mm}
\bigskip\hrule


\section{Introduction}\label{section1}

In conformal field theory (CFT), it is known that the general structure of three-point functions of conserved currents is highly constrained by conformal symmetry. 
A systematic approach to study three-point functions of primary operators was introduced in \cite{Osborn:1993cr, Erdmenger:1996yc} 
(see also 
refs.~\cite{Polyakov:1970xd, Schreier:1971um, Migdal:1971xh, Migdal:1971fof, Ferrara:1972cq, Ferrara:1973yt, Koller:1974ut, Mack:1976pa, Fradkin:1978pp, Stanev:1988ft} 
for earlier works), which presented an analysis of the general structure of three-point functions involving the energy-momentum 
tensor and conserved vector currents. The analysis of three-point functions of conserved higher-spin bosonic currents was later undertaken by 
Stanev~\cite{Stanev:2012nq, Stanev:2013eha, Stanev:2013qra} (see also~\cite{Elkhidir:2014woa,Buchbinder:2022cqp}) in the four-dimensional case, and by 
Zhiboedov \cite{Zhiboedov:2012bm} in general 
dimensions.\footnote{The study of correlation functions of conserved currents has also been extended to superconformal field theories in diverse 
dimensions \cite{Park:1997bq, Osborn:1998qu, Park:1998nra, Park:1999pd, Park:1999cw, Kuzenko:1999pi, Nizami:2013tpa, 
Buchbinder:2015qsa, Buchbinder:2015wia, Kuzenko:2016cmf, Buchbinder:2021gwu, Buchbinder:2021izb, Buchbinder:2021kjk, Buchbinder:2021qlb, 
Jain:2022izp, Buchbinder:2022kmj,Buchbinder:2023fqv,Buchbinder:2023ndg}.} 
In four dimensions (4D) it was shown that the number of independent structures in the three-point function of conserved bosonic vector-like 
currents $J_{\mu_1\dots \mu_s}$ increases linearly with the minimum spin. 
This is quite different to the results found in three dimensions (3D), where it has been shown by many 
authors \cite{Giombi:2011rz,Maldacena:2011jn,Giombi:2016zwa,Jain:2021whr,Jain:2021gwa,Jain:2021vrv,Buchbinder:2022mys} that there are only three possible
independent conserved structures for currents of arbitrary integer/half-integer spins. 
The aim of this paper is to study three-point functions of conserved currents belonging to arbitrary Lorentz representations in 4D CFT.  
An approach to this problem was outlined in \cite{Elkhidir:2014woa}, however, it did not study correlation functions when the operators are all conserved currents.

In this paper we provide a complete classification of three-point functions of conserved currents $J_{\a(i) \ad(j)}$, with $i,j \geq 1$ in four-dimensional conformal field theory.
Such currents satisfy the conservation equation
\begin{equation}
	\pa^{\b \bd} J_{\b \a(i-1) \bd \ad(j-1)} = 0 \, ,
\end{equation}
and possess scale dimension $ \D_{J} = s + 2 $, where the spin $s$ is given by $s = \frac{1}{2}(i+j)$. 
To classify the possible three-point functions of currents $J_{\a(i) \ad(j)}$, we find it more convenient to parametrise them in terms of their spin, $s$, and an integer, $q=i-j$, as follows:
\begin{align} \label{Current convention}
	J_{(s,q)} := J_{\a(s+\frac{q}{2}) \ad(s-\frac{q}{2})} \, .
\end{align}
With this convention $q$ is necessarily even/odd when $s$ is integer/half-integer valued. Note that the complex conjugate of $J_{(s,q)}$ is $J_{(s, -q)}$, hence, we introduce $\bar{J}_{(s,q)} := J_{(s, -q)}$ and view $q$ as being non-negative, taking values $q = 0, 1, ..., 2s - 2$. The case $q = 0$ corresponds to ``standard" bosonic conserved currents $J_{(s,0)} \equiv J_{\a(s) \ad(s)}$. Likewise, for $q=1$ we obtain pairs 
of (higher-spin) ``supersymmetry-like" currents $J_{(s,1)} \equiv J_{\a(s + \frac{1}{2}) \ad(s - \frac{1}{2})}$, $\bar{J}_{(s,1)} \equiv J_{(s,-1)} = J_{\a(s - \frac{1}{2}) \ad(s + \frac{1}{2})}$, where $s$ is necessarily half-integer valued. For example, by setting $s = \tfrac{3}{2}$ we obtain supersymmetry currents. In non-supersymmetric settings, the currents with $i=j$ (i.e. $q = 0$), were constructed explicitly in terms of free fields in  \cite{Craigie:1983fb}.

The main consequence of the notation \eqref{Current convention} for the currents is that there are essentially only two types of three-point functions to consider:
\begin{align} \label{possible three point functions}
	\langle J^{}_{(s_{1}, q_{1})}(x_{1}) \, J'_{(s_{2}, q_{2})}(x_{2}) \, J''_{(s_{3}, q_{3})}(x_{3}) \rangle \, , \hspace{10mm} \langle J^{}_{(s_{1}, q_{1})}(x_{1}) \, 
	\bar{J}'_{(s_{2}, q_{2})}(x_{2}) \, J''_{(s_{3}, q_{3})}(x_{3}) \rangle \, .
\end{align}
Any other possible three-point functions are equivalent to these up to permutations of the points or complex conjugation. 
The main aim of this paper is to develop a general formalism to study the structure of the three-point correlation functions \eqref{possible three point functions}, 
where we assume only the constraints imposed by conformal symmetry and conservation equations. In doing so we essentially provide a complete 
classification of all possible conserved three-point functions in 4D CFT. The three-point functions of currents with $q = 0, 1$ have been studied in 
e.g. \cite{Stanev:2012nq, Zhiboedov:2012bm, Elkhidir:2014woa, Buchbinder:2022cqp}. For bosonic conserved currents ($q_{i} = 0$), 
it is known that three-point functions of conserved currents with spins $s_{1}, s_{2}, s_{3}$ are fixed up to $2 \min(s_{1}, s_{2}, s_{3}) + 1$ solutions in general. We show that the same result also holds for three-point functions involving conserved currents with $q = 1$. The three-point functions of currents $J_{(s,q)}$ with $q \geq 2$, however, are relatively unexplored in the literature. Conserved currents with $q\geq2$ naturally arise in superconformal field theories in four-dimensions. As an example, consider a $\cN=2$ superconformal field theory possessing a conserved higher-spin supercurrent, $\mathbfcal{J}_{\a(s) \ad(s)}$ with $s \geq 1$, satisfying the following superfield conservation equation \cite{Howe:1981qj}:
\begin{equation}
	D_{i}^{\b} \mathbfcal{J}_{ \b \a(s-1) \ad(s)} = 0 \, ,
\end{equation}
where $D^{\b}_{i}$ is the spinor covariant derivative in $\cN=2$ superspace, and $i = 1,2$ is an iso-spinor index. The component structure of these supercurrents was elucidated in \cite{Kuzenko:2021pqm, Kuzenko:2023vgf}. The $\cN = 2$ supercurrent $\mathbfcal{J}_{ \a(s) \ad(s)}$ can be decomposed into the following collection of independent conformal $\cN=1$ supercurrent multiplets (see \cite{Kuzenko:2021pqm, Kuzenko:2023vgf} for more details):
\begin{align}
	\big\{ \mathbf{J}_{\a(s) \ad(s)},
	\mathbf{J}_{\a(s+1) \ad(s)},
	\mathbf{J}_{\a(s+1) \ad(s+1)} \big\} \, .
\end{align}
These $\cN=1$ supercurrents, in turn, contain a multiplet of conserved component currents \cite{Kuzenko:2019tys}. In particular, the $\cN=1$ supercurrent, $\mathbf{J}_{\a(s+1) \ad(s)}$,\footnote{This supercurrent was constructed explicitly in terms of a free massless hypermultiplet in \cite{Kuzenko:2017ujh}.} contains a conserved component current, $S_{\a(s+2) \ad(s)}$, defined as follows:
\begin{align}
	S_{\a(s+2) \ad(s)} &= D_{\a} \mathbf{J}_{\a(s+1) \ad(s)} | \, ,
\end{align}
where implicit symmetrisation among all $\a$-indices is assumed. Hence, the $\cN=2$ higher-spin supercurrent $\mathbfcal{J}_{\a(s) \ad(s)}$ contains 
a conserved component current $S_{\a(s+2) \ad(s)}$, which corresponds to $q=2$ in our convention 
above.
The $\cN=2$ supercurrents have been constructed explicitly for the free hypermultiplet and vector multiplet in \cite{Kuzenko:2021pqm, Kuzenko:2023vgf}.  

The formalism, which augments the approach of \cite{Osborn:1993cr} with auxiliary spinors, is suitable for constructing three-point functions of (conserved) primary operators in any Lorentz representation. Our approach is exhaustive in the sense that we construct all possible structures for the three-point function 
consistent with the conformal properties of the fields. We then systematically extract the linearly independent structures and impose the constraints arising from conservation equations, 
reality conditions, properties under inversion, and symmetries under permutations of spacetime points. The calculations are automated for arbitrary spins, and as a result we obtain the three-point function in an explicit form which can be presented up to our computational limit, $s_{i} = 10$. However, this limit is sufficient to propose a general classification of the results. 

We would like to point out that though the formalism developed in this paper is conceptually similar to the one developed for 
three-dimensional CFT in~\cite{Buchbinder:2022mys}, there are two important differences. First, in three dimensions, three-point functions of conserved 
currents can have at most three independent structures (two parity-even and one parity-odd), whereas in four dimensions the number of 
conserved structures (generically) grows linearly with the minimum spin. 
Second, for three-point functions in 3D CFT an important role is played by the triangle inequalities
\begin{align}
	s_{1} \leq s_{2} + s_{3} \, , && s_{2} \leq s_{1} + s_{3} \, , && s_{3} \leq s_{1} + s_{2} \, .
\end{align}
For three-point functions involving conserved currents which are within the triangle inequalities, there are two parity-even solutions and one parity-odd solution. 
However, if any of the triangle inequalities are not satisfied then the parity-odd solution is incompatible with conservation 
equations \cite{Giombi:2011rz,Maldacena:2011jn,Giombi:2016zwa,Jain:2021whr,Jain:2021gwa,Jain:2021vrv,Buchbinder:2022mys}. 
This statement has been proven in the light-cone limit in \cite{Maldacena:2011jn,Giombi:2016zwa} (see also \cite{Jain:2021whr} for results in 
momentum space). However, we found that in 4D CFT the triangle inequalities appear to have no significance.

The content of this paper is organised as follows. In section \ref{section2} we review the properties of the conformal building blocks used to construct correlation functions 
of primary operators in four dimensions. We then develop the formalism to construct three-point functions of primary operators of the form $J_{\a(i) \ad(j)}$, 
where we demonstrate how to impose all constraints arising from conservation equations, reality conditions and point switch symmetries. In particular, we utilise an index-free auxiliary spinor formalism 
to construct a generating function for the three-point functions, and we outline the pertinent aspects of our computational approach. In section \ref{section3}, we demonstrate 
our formalism by analysing the structure of three-point functions involving conserved vector currents, ``supersymmetry-like" currents and the energy-momentum tensor. 
We reproduce the known results previously found in~\cite{Osborn:1993cr,Stanev:2012nq,Zhiboedov:2012bm}. We then expand our discussion to include three-point functions of 
higher-spin currents belonging to any Lorentz representation, and provide a classification of the results.
For this the structure of the solutions is more easily identified by using the notation 
$J^{}_{(s,q)},\  \bar{J}_{(s,q)}$ for the currents as outlined above. In particular, we show that special attention is required for three-point functions of the form 
$\langle J^{}_{(s_{1},q)} \bar{J}'_{(s_{2},q)} J''_{(s_{3},0)} \rangle$ with $q \geq 2$. In this case the formula for the number of independent conserved structures is 
found to be quite non-trivial. The appendix \ref{AppA} is devoted to mathematical conventions and various useful identities. In appendix \ref{AppB} we 
provide some examples of the three-point functions $\langle J^{}_{(s_{1},q)} \bar{J}'_{(s_{2},q)} J''_{(s_{3},0)} \rangle$ for which the number of independent 
conserved structures differs from $2 \min(s_{1}, s_{2}, s_{3}) + 1$.
Then, as a consistency check, in appendix \ref{AppC} we provide further examples of three-point functions involving scalars, spinors and conserved currents to compare against the results in \cite{Elkhidir:2014woa}.


\section{Conformal building blocks}\label{section2}

In this section we will review the group theoretic formalism used to compute correlation functions of primary operators in four dimensional conformal field theories. For a more detailed introduction to the formalism as applied to correlation functions of bosonic primary fields see \cite{Osborn:1993cr}. Our 4D conventions and notation are outlined in appendix \ref{AppA}.

\subsection{Two-point functions}\label{section2.1}

Consider 4D Minkowski space $\mathbb{M}^{1,3}$, parameterised by coordinates $ x^{m} $, where $m = 0, 1, 2, 3$ are Lorentz indices. For any two points, $x_{1}, x_{2}$, we construct the covariant two-point functions 
\begin{align}
	x_{12 \, m} &= (x_{1} - x_{2})_{m} \, .
\end{align}
The two-point functions can be converted to spinor notation using the conventions outlined in appendix \ref{AppA}:
\begin{align}
	x_{12 \, \a \ad} &= (\s^{m})_{\a \ad} x_{12 \, m} \, , & x_{12}^{\ad \a} &= (\tilde{\s}^{m})^{\ad \a} x_{12 \, m} \, , & x_{12}^{2} &= - \frac{1}{2} x_{12}^{\ad \a} x^{}_{12 \, \a \ad} \, .
\end{align}
In this form the two-point functions possess the following useful properties:
\begin{align}  \label{Two-point building blocks - properties 1} 
	x_{12}^{\ad \a} x^{}_{12 \, \b \ad} = - x_{12}^{2} \d_{\b}^{\a} \, , \hspace{10mm} x_{12}^{\ad \a} x_{12 \, \a \bd} = - x_{12}^{2} \d_{\bd}^{\ad} \, . 
\end{align}
Hence, we find
\begin{equation} \label{Two-point building blocks 4}
	(x_{12}^{-1})^{\ad \a} = - \frac{x_{12}^{\ad \a}}{x_{12}^{2}} \, .
\end{equation}
We also introduce the normalised two-point functions, denoted by $\hat{x}_{12}$,
\begin{align} \label{Two-point building blocks 3}
	\hat{x}_{12 \, \a \ad} = \frac{x_{12 \, \a \ad}}{( x_{12}^{2})^{1/2}} \, , \hspace{10mm} \hat{x}_{12}^{\ad \a} \hat{x}^{}_{12 \, \b \ad} = - \d_{\b}^{\a} \, . 
\end{align}
From here we can now construct an operator analogous to the conformal inversion tensor acting on the space of symmetric traceless tensors of arbitrary rank. Given a two-point function, $x$, we define the operator
\begin{equation} \label{Higher-spin inversion operators a}
	\cI_{\a(k) \ad(k)}(x) = \hat{x}_{(\a_{1} (\ad_{1}} \dots \hat{x}_{ \a_{k}) \ad_{k})}  \, ,
\end{equation}
along with its inverse
\begin{equation} \label{Higher-spin inversion operators b}
	\bar{\cI}^{\ad(k) \a(k)}(x) = \hat{x}^{(\ad_{1} (\a_{1}} \dots \hat{x}^{ \ad_{k}) \a_{k})} \, .
\end{equation}
The spinor indices may be raised and lowered using the standard conventions as follows:
\begin{subequations}
	\begin{align}
		\cI^{\a(k)}{}_{\ad(k)}(x) &= \ve^{\a_{1} \g_{1}} \dots \ve^{\a_{k} \g_{k}} \, \cI_{\g(k) \ad(k)}(x) \, , \\
		\bar{\cI}_{\ad(k)}{}^{\a(k)}(x) &= \ve_{\ad_{1} \gd_{1}} \dots \ve_{\ad_{k} \gd_{k}} \, \bar{\cI}^{\gd(k) \a(k)}(x) \, .
	\end{align}
\end{subequations}
Now due to the property
\begin{equation}
	\cI_{\a(k) \ad(k)}(-x) = (-1)^{k} \cI_{\a(k) \ad(k)}(x) \, ,
\end{equation}
we have the following useful relations:
\begin{subequations} \label{Higher-spin inversion operators - properties}
	\begin{align}
		\cI_{\a(k) \ad(k)}(x_{12}) \, \bar{\cI}^{\ad(k) \b(k)}(x_{21}) &= \d_{(\a_{1}}^{(\b_{1}} \dots \d_{\a_{k})}^{\b_{k})} \, , \\
		\bar{\cI}^{\bd(k) \a(k)}(x_{12}) \, \cI_{\a(k) \ad(k)}(x_{21}) &= \d_{(\ad_{1}}^{(\bd_{1}} \dots \d_{\ad_{k})}^{\bd_{k})} \, .
	\end{align}
\end{subequations}
The objects \eqref{Higher-spin inversion operators a}, \eqref{Higher-spin inversion operators b} prove to be essential in the construction of correlation functions involving primary operators of arbitrary spins. Indeed, the vector representation of the inversion tensor may be recovered in terms of the spinor two-point functions as follows:
\begin{equation}
	I_{m n}(x) = - \frac{1}{2} \, \text{Tr}( \tilde{\s}_{m} \, \hat{x} \, \tilde{\s}_{n} \, \hat{x} ) \, .
\end{equation}
%
%
Now let $\F_{\cA}$ be a primary field with dimension $\D$, where $\cA$ denotes a collection of Lorentz spinor indices. The two-point correlation function of $\F_{\cA}$ and its conjugate $\bar{\F}^{\bar{\cA}}$ is fixed by conformal symmetry to the form
\begin{equation}
	\langle \F_{\cA}(x_{1}) \, \bar{\F}^{\bar{\cA}}(x_{2}) \rangle = c \, \frac{\cI_{\cA}{}^{\bar{\cA}}(x_{12})}{(x_{12}^{2})^{\D}} \, , 
\end{equation} 
where $\cI$ is an appropriate representation of the inversion tensor and $c$ is a constant complex parameter. The denominator of the two-point function is determined by the conformal dimension of $\F_{\cA}$, which guarantees that the correlation function transforms with the appropriate weight under scale transformations.


\subsection{Three-point functions}\label{section2.2}

Given three distinct points in Minkowski space, $x_{i}$, with $i = 1,2,3$, we define conformally covariant three-point functions in terms of the two-point functions as in \cite{Osborn:1993cr}
\begin{align}
	X_{ij} &= \frac{x_{ik}}{x_{ik}^{2}} - \frac{x_{jk}}{x_{jk}^{2}} \, , & X_{ji} &= - X_{ij} \, ,  & X_{ij}^{2} &= \frac{x_{ij}^{2}}{x_{ik}^{2} x_{jk}^{2} } \, , 
\end{align}
where $(i,j,k)$ is a cyclic permutation of $(1,2,3)$. For example, we have
\begin{equation}
	X_{12}^{m} = \frac{x_{13}^{m}}{x_{13}^{2}} - \frac{x_{23}^{m}}{x_{23}^{2}} \, , \hspace{10mm} X_{12}^{2} = \frac{x_{12}^{2}}{x_{13}^{2} x_{23}^{2} } \, .
\end{equation}
There are several useful identities involving the two- and three-point functions and the conformal inversion tensor. For example we have the useful algebraic relations
\begin{subequations}
	\begin{align} \label{Inversion tensor identities - vector case 1}
		I_{m}{}^{a}(x_{13}) \, I_{a n}(x_{23}) &= I_{m}{}^{a}(x_{12}) \, I_{a n}(X_{13}) \, ,  & I_{m n}(x_{23}) \, X_{12}^{n} &= \frac{x_{12}^{2}}{x_{13}^{2}} \, X_{13 \, m} \, ,
	\end{align} \vspace{-6mm}
	\begin{align} \label{Inversion tensor identities - vector case 2}
		I_{m}{}^{a}(x_{23}) \, I_{a n}(x_{13}) &= I_{m}{}^{a}(x_{21}) \, I_{a n}(X_{32}) \, ,  & I_{m n}(x_{13}) \, X_{12}^{n} &= \frac{x_{12}^{2}}{x_{23}^{2}} \, X_{32 \, m} \, , 
	\end{align}
\end{subequations}
and the differential identities
\begin{align}
	\pa^{(1)}_{m} X_{12 \, n} = \frac{1}{x_{13}^{2}} I_{m n}(x_{13}) \, , \hspace{10mm} \pa^{(2)}_{m} X_{12 \, n} = - \frac{1}{x_{23}^{2}} I_{m n}(x_{23}) \, . \label{Inversion tensor identities - vector case 3}
\end{align}
The three-point functions also may be represented in spinor notation as follows:
\begin{equation}
	X_{ij , \, \a \ad} = (\s_{m})_{\a \ad} X_{ij}^{m} \, , \hspace{10mm} X_{ij , \, \a \ad} = (x^{-1}_{ik})_{\a \gd} x_{ij}^{\gd \g} (x^{-1}_{jk})_{\g \ad} \, .
\end{equation}
These objects satisfy properties similar to the two-point functions \eqref{Two-point building blocks - properties 1}. Indeed, 
it is convenient to define the normalised three-point functions, $\hat{X}_{ij}$, and the inverses, $(X_{ij}^{-1})$
\begin{equation}
	\hat{X}_{ij , \, \a \ad} = \frac{X_{ij , \, \a \ad}}{( X_{ij}^{2})^{1/2}} \, , \hspace{10mm}	(X_{ij}^{-1})^{\ad \a} = - \frac{X_{ij}^{\ad \a}}{X_{ij}^{2}} \, .
\end{equation}  
Now given an arbitrary three-point building block $X$, it is also useful to construct the following higher-spin operator:
\begin{equation}
	\cI_{\a(k) \ad(k)}(X) = \hat{X}_{ (\a_{1} (\ad_{1}} \dots \hat{X}_{\a_{k}) \ad_{k})}  \, ,
\end{equation}
along with its inverse
\begin{equation}
	\bar{\cI}^{\ad(k) \a(k)}(X) = \hat{X}^{(\ad_{1} (\a_{1}} \dots \hat{X}^{ \ad_{k}) \a_{k})} \, .
\end{equation}
These operators have properties similar to the two-point higher-spin inversion operators \eqref{Higher-spin inversion operators a}, \eqref{Higher-spin inversion operators b}. There are also some useful algebraic identities relating the two- and three-point functions at various points, such as
\begin{equation}
	\cI_{\a \ad}(X_{12})  = \cI_{\a \gd}(x_{13}) \, \bar{\cI}^{\gd \g}(x_{12}) \, \cI_{\g \ad}(x_{23}) \, , \hspace{5mm}  \bar{\cI}^{\ad \g}(x_{13}) \, \cI_{\g \gd}(X_{12})  \, \bar{\cI}^{\gd \a}(x_{13}) = \bar{\cI}^{\ad \a}(X_{32})  \, . \label{Inversion tensor identities - spinor case}
\end{equation}
These identities are analogous to \eqref{Inversion tensor identities - vector case 1}, \eqref{Inversion tensor identities - vector case 2}, and admit generalisations to higher-spins, for example
\begin{equation}
	\bar{\cI}^{\ad(k) \g(k)}(x_{13}) \, \cI_{\g(k) \gd(k)}(X_{12}) \, \bar{\cI}^{\gd(k) \a(k)}(x_{13}) = \bar{\cI}^{\ad(k) \a(k)}(X_{32})  \, . \label{Inversion tensor identities - higher spin case}
\end{equation}
In addition, similar to \eqref{Inversion tensor identities - vector case 3}, there are also the following useful identities:
\begin{equation}
	\pa^{(1)}_{\a \ad} X_{12}^{ \dot{\s} \s} = - \frac{2}{x_{13}^{2}} \, \cI_{\a}{}^{\dot{\s}}(x_{13}) \,  \bar{\cI}_{\ad}{}^{\s}(x_{13}) \, , \hspace{5mm} \pa^{(2)}_{\a \ad} X_{12}^{ \dot{\s} \s} = \frac{2}{x_{23}^{2}} \, \cI_{\a}{}^{\dot{\s}}(x_{23}) \,  \bar{\cI}_{\ad}{}^{\s}(x_{23}) \, . \label{Three-point building blocks - differential identities}
\end{equation}
These identities allow us to account for the fact that correlation functions of primary fields can obey differential constraints which can arise due to conservation equations. Indeed, given a tensor field $\cT_{\cA}(X)$, there are the following differential identities which arise as a consequence of \eqref{Three-point building blocks - differential identities}:
\begin{subequations}
	\begin{align}
		\pa_{(1) \, \a \ad} \cT_{\cA}(X_{12}) &= \frac{1}{x_{13}^{2}} \, \cI_{\a}{}^{\dot{\s}}(x_{13}) \,  \bar{\cI}_{\ad}{}^{\s}(x_{13}) \, \frac{ \pa}{ \pa X_{12}^{ \dot{\s} \s}} \, \cT_{\cA}(X_{12}) \, ,  \label{Three-point building blocks - differential identities 2} \\[2mm]
		\pa_{(2) \, \a \ad} \cT_{\cA}(X_{12}) &= - \frac{1}{x_{23}^{2}} \, \cI_{\a}{}^{\dot{\s}}(x_{23}) \,  \bar{\cI}_{\ad}{}^{\s}(x_{23}) \, \frac{ \pa}{ \pa X_{12}^{ \dot{\s} \s}} \, \cT_{\cA}(X_{12}) \, . \label{Three-point building blocks - differential identities 3}
	\end{align}
\end{subequations}
Now let $\F$, $\J$, $\P$ be primary fields with scale dimensions $\D_{1}$, $\D_{2}$ and $\D_{3}$ respectively. The three-point function may be constructed using the general ansatz
\begin{align}
	\langle \F_{\cA_{1}}(x_{1}) \, \J_{\cA_{2}}(x_{2}) \, \P_{\cA_{3}}(x_{3}) \rangle = \frac{ \cI^{(1)}{}_{\cA_{1}}{}^{\bar{\cA}_{1}}(x_{13}) \,  \cI^{(2)}{}_{\cA_{2}}{}^{\bar{\cA}_{2}}(x_{23}) }{(x_{13}^{2})^{\D_{1}} (x_{23}^{2})^{\D_{2}}}
	\; \cH_{\bar{\cA}_{1} \bar{\cA}_{2} \cA_{3}}(X_{12}) \, , \label{Three-point function - general ansatz}
\end{align} 
where the tensor $\cH_{\bar{\cA}_{1} \bar{\cA}_{2} \cA_{3}}$ encodes all information about the correlation function, and is constrained by the conformal symmetry as follows:
\begin{enumerate}
	\item[\textbf{(i)}] Under scale transformations of Minkowski space $x^{m} \mapsto x'^{m} = \l^{-2} x^{m}$, the three-point building blocks transform as $X^{m} \mapsto X'^{m} = \l^{2} X^{m}$. As a consequence, the correlation function transforms as 
	\begin{equation}
		\langle \F_{\cA_{1}}(x_{1}') \, \J_{\cA_{2}}(x_{2}') \, \P_{\cA_{3}}(x_{3}') \rangle = (\l^{2})^{\D_{1} + \D_{2} + \D_{3}} \langle \F_{\cA_{1}}(x_{1}) \, \J_{\cA_{2}}(x_{2}) \,  \P_{\cA_{3}}(x_{3}) \rangle \, ,
	\end{equation}
	which implies that $\cH$ obeys the scaling property
	\begin{equation}
		\cH_{\bar{\cA}_{1} \bar{\cA}_{2} \cA_{3}}(\l^{2} X) = (\l^{2})^{\D_{3} - \D_{2} - \D_{1}} \, \cH_{\bar{\cA}_{1} \bar{\cA}_{2} \cA_{3}}(X) \, , \hspace{5mm} \forall \l \in \mathbb{R} \, \backslash \, \{ 0 \} \, .
	\end{equation}
	This guarantees that the correlation function transforms correctly under scale transformations.
	
	\item[\textbf{(ii)}] If any of the fields $\F$, $\J$, $\P$ obey differential equations, such as conservation equations, then the tensor $\cH$ is also constrained by differential equations. Such constraints may be derived with the aid of identities \eqref{Three-point building blocks - differential identities 2}, \eqref{Three-point building blocks - differential identities 3}.
	
	\item[\textbf{(iii)}] If any (or all) of the operators $\F$, $\J$, $\P$ coincide, the correlation function possesses symmetries under permutations of spacetime points, e.g.
	\begin{equation}
		\langle \F_{\cA_{1}}(x_{1}) \, \F_{\cA'_{1}}(x_{2}) \, \P_{\cA_{3}}(x_{3}) \rangle = (-1)^{\e(\F)} \langle \F_{\cA'_{1}}(x_{2}) \, \F_{\cA_{1}}(x_{1}) \, \P_{\cA_{3}}(x_{3}) \rangle \, ,
	\end{equation}
	where $\e(\F)$ is the Grassmann parity of $\F$. As a consequence, the tensor $\cH$ obeys constraints which will be referred to as ``point-switch symmetries". A similar relation may also be derived for two fields which are related by complex conjugation.

\end{enumerate}

The constraints above fix the functional form of $\cH$ (and therefore the correlation function) up to finitely many independent parameters. Hence, using the general formula \eqref{H ansatz}, the problem of computing three-point correlation functions is reduced to deriving the general structure of the tensor $\cH$ subject to the above constraints.	

\subsubsection{Comments on differential constraints}\label{subsubsection2.2.1}

For three-point functions of conserved currents, we must impose conservation on all three space-time points. For $x_{1}$ and $x_{2}$, this process is simple due to the identities \eqref{Three-point building blocks - differential identities 2}, \eqref{Three-point building blocks - differential identities 3}, and the resulting conservation equations become equivalent to simple differential constraints on $\cH$. However, conservation on $x_{3}$ is more challenging due to a lack of useful identities analogous to \eqref{Three-point building blocks - differential identities 2}, \eqref{Three-point building blocks - differential identities 3} for $x_{3}$. To correctly impose conservation on $x_{3}$, consider the correlation function $\langle \F_{\cA_{1}}(x_{1}) \, \J_{\cA_{2}}(x_{2}) \, \P_{\cA_{3}}(x_{3}) \rangle$, with the ansatz
\begin{equation} \label{H ansatz}
	\langle \F_{\cA_{1}}(x_{1}) \, \J_{\cA_{2}}(x_{2}) \, \P_{\cA_{3}}(x_{3}) \rangle = \frac{ \cI^{(1)}{}_{\cA_{1}}{}^{\bar{\cA}_{1}}(x_{13}) \,  \cI^{(2)}{}_{\cA_{2}}{}^{\bar{\cA}_{2}}(x_{23}) }{(x_{13}^{2})^{\D_{1}} (x_{23}^{2})^{\D_{2}}}
	\; \cH_{\bar{\cA}_{1} \bar{\cA}_{2} \cA_{3}}(X_{12}) \, . 
\end{equation} 
We now reformulate the ansatz with $\P$ at the front
\begin{equation} \label{Htilde ansatz}
	\langle \P_{\cA_{3}}(x_{3}) \, \J_{\cA_{2}}(x_{2}) \, \F_{\cA_{1}}(x_{1}) \rangle = \frac{ \cI^{(3)}{}_{\cA_{3}}{}^{\bar{\cA}_{3}}(x_{31}) \,  \cI^{(2)}{}_{\cA_{2}}{}^{\bar{\cA}_{2}}(x_{21}) }{(x_{31}^{2})^{\D_{3}} (x_{21}^{2})^{\D_{2}}}
	\; \tilde{\cH}_{\bar{\cA}_{3} \bar{\cA}_{2} \cA_{1}}(X_{32}) \, . 
\end{equation} 
These two correlators are the same up to an overall sign due to Grassmann parity. Equating the two ansatz above yields the following relation:
\begin{align} \label{Htilde and H relation}
	\begin{split}
		\tilde{\cH}_{\bar{\cA}_{3} \bar{\cA}_{2} \cA_{1}}(X_{32}) &= (x_{13}^{2})^{\D_{3} - \D_{1}} \bigg(\frac{x_{21}^{2}}{x_{23}^{2}} \bigg)^{\hspace{-1mm} \D_{2}} \, \cI^{(1)}{}_{\cA_{1}}{}^{\bar{\cA}_{1}}(x_{13}) \, \bar{\cI}^{(2)}{}_{\bar{\cA}_{2}}{}^{\cA'_{2}}(x_{12}) \,  \cI^{(2)}{}_{\cA'_{2}}{}^{\bar{\cA}'_{2}}(x_{23}) \\[-2mm]
		& \hspace{50mm} \times \bar{\cI}^{(3)}{}_{\bar{\cA}_{3}}{}^{\cA_{3}}(x_{13}) \, \cH_{\bar{\cA}_{1} \bar{\cA}'_{2} \cA_{3}}(X_{12}) \, . 
	\end{split}
\end{align}
Now suppose $\cH(X)$ (with indices suppressed) is composed of finitely many linearly independent tensor structures, $P_{i}(X)$, i.e $\cH(X) = \sum_{i} a_{i} P_{i}(X)$ where $a_{i}$ are constant complex parameters. We define $\bar{\cH}(X) = \sum_{i} \bar{a}_{i} \bar{P}_{i}(X)$, the conjugate of $\cH$, and also $\cH^{c}(X) = \sum_{i} a_{i} \bar{P}_{i}(X)$, which we denote as the complement of $\cH$. As a consequence of \eqref{Inversion tensor identities - spinor case}, the following relation holds:
\begin{align} \label{Hc and H relation}
	\begin{split}
		\cH^{c}_{\cA_{1} \cA_{2} \bar{\cA}_{3}}(X_{32}) &= (x_{13}^{2} X_{32}^{2})^{\D_{3} - \D_{2} - \D_{1}} \cI^{(1)}{}_{\cA_{1}}{}^{\bar{\cA}_{1}}(x_{13}) \, \cI^{(2)}{}_{\cA_{2}}{}^{\bar{\cA}_{2}}(x_{13}) \\
		& \hspace{45mm} \times \bar{\cI}^{(3)}{}_{\bar{\cA}_{3}}{}^{\cA_{3}}(x_{13}) \, \cH_{\bar{\cA}_{1} \bar{\cA}_{2} \cA_{3}}(X_{12}) \, .
	\end{split}
\end{align}
After inverting this identity and substituting it directly into \eqref{Htilde and H relation}, we apply \eqref{Inversion tensor identities - spinor case} to obtain an equation relating $\cH^{c}$ and $\tilde{\cH}$
\begin{equation} \label{Htilde and Hc relation}
	\tilde{\cH}_{\bar{\cA}_{3} \bar{\cA}_{2} \cA_{1}}(X) = (X^{2})^{\D_{1} - \D_{3}} \, \bar{\cI}^{(2)}{}_{\bar{\cA}_{2}}{}^{\cA_{2}}(X) \, \cH^{c}_{\cA_{1} \cA_{2} \bar{\cA}_{3}}(X) \, . 
\end{equation}
Conservation on $x_{3}$ may now be imposed by using \eqref{Three-point building blocks - differential identities 2}, with $ x_{1} \leftrightarrow x_{3}$. In principle, this procedure can be carried out for any configuration of the fields.

If we now consider the correlation function of three conserved primaries $J^{}_{\a(i_{1}) \ad(j_{1})}$, $J'_{\b(i_{2}) \bd(j_{2})}$, $J''_{\g(i_{3}) \gd(j_{3})}$, where $s_{1} = \tfrac{1}{2}(i_{1} + j_{1})$, $s_{2} = \tfrac{1}{2}( i_{2} + j_{2} )$, $s_{3} = \tfrac{1}{2}( i_{3} + j_{3} )$, then the general ansatz is
\begin{align} \label{Conserved correlator ansatz}
	\langle \, J^{}_{\a(i_{1}) \ad(j_{1})}(x_{1}) \, J'_{\b(i_{2}) \bd(j_{2})}(x_{2}) \, J''_{\g(i_{3}) \gd(j_{3})}(x_{3}) \rangle &= \nonumber \\
	& \hspace{-55mm} \frac{1}{ (x_{13}^{2})^{\D_{1}} (x_{23}^{2})^{\D_{2}} } \, \cI_{\a(i_{1})}{}^{\ad'(i_{1})}(x_{13}) \, \bar{\cI}_{\ad(j_{1})}{}^{\a'(j_{1})}(x_{13}) \,  \cI_{\b(i_{2})}{}^{\bd'(i_{2})}(x_{23}) \, \bar{\cI}_{\bd(j_{2})}{}^{\b'(j_{2})}(x_{23}) \nonumber\\
	& \hspace{-25mm} \times \cH_{\a'(j_{1}) \ad'(i_{1}) \b'(j_{2}) \bd'(i_{2}) \g(i_{3}) \gd(j_{3})}(X_{12}) \, ,
\end{align} 
where $\D_{i} = s_{i} + 2$. The constraints on $\cH$ are then as follows:
\begin{enumerate}
	\item[\textbf{(i)}] {\bf Homogeneity:} \\
	Recall that $\cH$ is a homogeneous tensor field satisfying
	\begin{equation}
		\cH_{\a(j_{1}) \ad(i_{1}) \b(j_{2}) \bd(i_{2}) \g(i_{3}) \gd(j_{3})}(\l^{2} X) = (\l^{2})^{\D_{3} - \D_{2} - \D_{1}} \, \cH_{\a(j_{1}) \ad(i_{1}) \b(j_{2}) \bd(i_{2}) \g(i_{3}) \gd(j_{3})}(X) \, .
	\end{equation}
	It is often convenient to introduce $\hat{\cH}_{\a(j_{1}) \ad(i_{1}) \b(j_{2}) \bd(i_{2}) \g(i_{3}) \gd(j_{3})}(X)$, such that
	\begin{align}
		\cH_{\a(j_{1}) \ad(i_{1}) \b(j_{2}) \bd(i_{2}) \g(i_{3}) \gd(j_{3})}(X) &= X^{\D_{3} - \D_{2}- \D_{1}} \hat{\cH}_{\a(j_{1}) \ad(i_{1}) \b(j_{2}) \bd(i_{2}) \g(i_{3}) \gd(j_{3})}(X) \, ,
	\end{align}
	where $\hat{\cH}_{\a(j_{1}) \ad(i_{1}) \b(j_{2}) \bd(i_{2}) \g(i_{3}) \gd(j_{3})}(X)$ is homogeneous degree 0 in $X$, i.e.
	\begin{align}
		\hat{\cH}_{\a(j_{1}) \ad(i_{1}) \b(j_{2}) \bd(i_{2}) \g(i_{3}) \gd(j_{3})}( \l^{2} X) &= \hat{\cH}_{\a(j_{1}) \ad(i_{1}) \b(j_{2}) \bd(i_{2}) \g(i_{3}) \gd(j_{3})}(X) \, .
	\end{align}
	
	\item[\textbf{(ii)}] {\bf Differential constraints:} \\
	After application of the identities \eqref{Three-point building blocks - differential identities 2}, \eqref{Three-point building blocks - differential identities 3} we obtain the following constraints:
	\begin{subequations}
		\begin{align}
			\text{Conservation at $x_{1}$:} && \pa_{X}^{\a \ad} \cH_{\a \a(j_{1} -1) \ad \ad(i_{1} -1) \b(j_{2}) \bd(i_{2}) \g(i_{3}) \gd(j_{3})}(X) &= 0 \, , \\
			\text{Conservation at $x_{2}$:} && \pa_{X}^{\b \bd} \cH_{\a(j_{1}) \ad(i_{1}) \b \b(j_{2}-1) \bd \bd(i_{2}-1) \g(i_{3}) \gd(j_{3})}(X) &= 0 \, , \\
			\text{Conservation at $x_{3}$:} && \pa_{X}^{\g \gd} \tilde{\cH}_{\a(i_{1}) \ad(j_{1}) \b(j_{2}) \bd(i_{2}) \g \g(j_{3}-1) \gd \gd(i_{3}-1)}(X) &= 0 \, ,
		\end{align}
	\end{subequations}
	where
	\begin{align}
		\begin{split}
			\tilde{\cH}_{\a(i_{1}) \ad(j_{1}) \b(j_{2}) \bd(i_{2}) \g(j_{3}) \gd(i_{3})}(X) &= (X^{2})^{\D_{1} - \D_{3}} \, \cI_{\b(j_{2})}{}^{\bd'(j_{2})}(X) \, \bar{\cI}_{\bd(i_{2})}{}^{\b'(i_{2})}(X) \\
			& \hspace{15mm}\times \cH^{c}_{\a(i_{1}) \ad(j_{1}) \b'(i_{2}) \bd'(j_{2}) \g(j_{3}) \gd(i_{3})}(X) \, . 
		\end{split}
	\end{align}

	\item[\textbf{(iii)}] {\bf Point-switch symmetries:} \\
	If the fields $J$ and $J'$ coincide, then we obtain the following point-switch identity
	\begin{equation}
		\cH_{\a(i_{1}) \ad(j_{1}) \b(i_{1}) \bd(j_{1}) \g(i_{3}) \gd(j_{3})}(X) = (-1)^{\e(J)} \cH_{\a(i_{1}) \ad(j_{1}) \b(i_{1}) \bd(j_{1}) \g(i_{3}) \gd(j_{3})}(-X) \, ,
	\end{equation}
	where $\e(J)$ is the Grassmann parity of $J$. Likewise, if the fields $J$ and $J''$ coincide, then we obtain the constraint
	\begin{equation}
	\tilde{\cH}_{\a(i_{1}) \ad(j_{1}) \b(j_{2}) \bd(i_{2}) \g(j_{1}) \gd(i_{1})}(X) = (-1)^{\e(J)} \cH_{\a(j_{1}) \ad(i_{1}) \b(j_{2}) \bd(i_{2}) \g(i_{1}) \gd(j_{1})}(-X)  \, .
	\end{equation}

	\item[\textbf{(iv)}] {\bf Reality condition:} \\
	If the fields in the correlation function belong to the $(s,s)$ representation, then the three-point function must satisfy the reality condition
	\begin{equation}
		\cH_{\a(i_{1}) \ad(i_{1}) \b(i_{2}) \bd(i_{2}) \g(i_{3}) \gd(i_{3})}(X) = \bar{\cH}_{\a(i_{1}) \ad(i_{1}) \b(i_{2}) \bd(i_{2}) \g(i_{3}) \gd(i_{3})}(X)  \, .
	\end{equation}
	Similarly, if the fields at $J$, $J'$ at $x_{1}$ and $x_{2}$ respectively possess the same spin and are conjugate to each other, i.e. $J' = \bar{J}$, we must impose a combined reality/point-switch condition using the following constraint
	\begin{equation}
		\cH_{\a(i_{1}) \ad(j_{1}) \b(j_{1}) \bd(i_{1}) \g(i_{3}) \gd(j_{3})}(X) = (-1)^{\e(J)} \bar{\cH}_{\b(i_{1}) \bd(j_{1}) \a(j_{1}) \ad(i_{1})  \g(i_{3}) \gd(j_{3})}(-X) \, ,
	\end{equation}
	where $\e(J)$ is the Grassmann parity of $J$. 
	
\end{enumerate}
Working with the tensor formalism is quite messy and complicated in general, hence, to simplify the analysis we will utilise auxiliary spinors to carry out the computations.

\subsubsection{Generating function formalism}\label{subsubsection2.2.2}

Analogous to the approach of \cite{Buchbinder:2022mys} we utilise auxiliary spinors to streamline the calculations. Consider a general spin-tensor $\cH_{\cA_{1} \cA_{2} \cA_{3}}(X)$, where $\cA_{1} = \{ \a(i_{1}), \ad(j_{1}) \}, \cA_{2} = \{ \b(i_{2}), \bd(j_{2}) \}, \cA_{3} = \{ \g(i_{3}), \gd(j_{3}) \}$ represent sets of totally symmetric spinor indices associated with the fields at points $x_{1}$, $x_{2}$ and $x_{3}$ respectively. We introduce sets of commuting auxiliary spinors for each point; $ U = \{ u, \bar{u} \}$ at $x_{1}$, $ V = \{ v, \bar{v} \}$ at $x_{2}$, and $W = \{ w, \bar{w} \}$ at $x_{3}$, where the spinors satisfy 
\begin{align}
	u^2 &= \varepsilon_{\a \b} \, u^{\a} u^{\b}=0\,, & \bar{u}^2& = \varepsilon_{\ad \bd} \, \bar{u}^{\ad} \bar{u}^{\bd}=0\,,  &
	v^{2} &= \bar{v}^{2} = 0\,, & w^{2} &= \bar{w}^{2} = 0\,. 
	\label{extra1}
\end{align}
Now if we define the objects
\begin{subequations}
	\begin{align}
		\boldsymbol{U}^{\cA_{1}} &\equiv \boldsymbol{U}^{\a(i_{1}) \ad(j_{1})} = u^{\a_{1}} \dots u^{\a_{i_{1}}} \bar{u}^{\ad_{1}} \dots \bar{u}^{\ad_{j_{1}}} \, , \\
		\boldsymbol{V}^{\cA_{2}} &\equiv \boldsymbol{V}^{\b(i_{2}) \bd(j_{2})} = v^{\b_{1}} \dots v^{\b_{i_{2}}} \bar{v}^{\bd_{1}} \dots \bar{v}^{\bd_{j_{2}}} \, , \\
		\boldsymbol{W}^{\cA_{3}} &\equiv \boldsymbol{W}^{\g(i_{3}) \gd(j_{3})} = w^{\g_{1}} \dots w^{\g_{i_{3}}} \bar{w}^{\gd_{1}} \dots \bar{w}^{\gd_{j_{3}}} \, ,
	\end{align}
\end{subequations}
then the generating polynomial for $\cH$ is constructed as follows:
\begin{equation} \label{H - generating polynomial}
	\cH(X; U, V, W) = \cH_{ \cA_{1} \cA_{2} \cA_{3} }(X) \, \boldsymbol{U}^{\cA_{1}} \boldsymbol{V}^{\cA_{2}} \boldsymbol{W}^{\cA_{3}} \, . \\
\end{equation}
The tensor $\cH$ can then be extracted from the polynomial by acting on it with the following partial derivative operators:
\begin{subequations}
	\begin{align}
		\frac{\pa}{\pa \boldsymbol{U}^{\cA_{1}} } &\equiv \frac{\pa}{\pa \boldsymbol{U}^{\a(i_{1}) \ad(j_{1})} } = \frac{1}{i_{1}!j_{1}!} \frac{\pa}{\pa u^{\a_{1}} } \dots \frac{\pa}{\pa u^{\a_{i_{1}}}} \frac{\pa}{\pa \bar{u}^{\ad_{1}}} \dots \frac{\pa }{\pa \bar{u}^{\ad_{j_{1}}}} \, , \\
		\frac{\pa}{\pa \boldsymbol{V}^{\cA_{2}} } &\equiv \frac{\pa}{\pa \boldsymbol{V}^{\b(i_{2}) \bd(j_{2})} } = \frac{1}{i_{2}!j_{2}!} \frac{\pa}{\pa v^{\b_{1}} } \dots \frac{\pa}{\pa v^{\b_{i_{2}}}} \frac{\pa}{\pa \bar{v}^{\bd_{1}}} \dots \frac{\pa }{\pa \bar{v}^{\bd_{j_{2}}}} \, , \\
		\frac{\pa}{\pa \boldsymbol{W}^{\cA_{3}} } &\equiv \frac{\pa}{\pa \boldsymbol{W}^{\g(i_{3}) \gd(j_{3})} } = \frac{1}{i_{3}!j_{3}!} \frac{\pa}{\pa w^{\g_{1}} } \dots \frac{\pa}{\pa w^{\g_{i_{3}}}} \frac{\pa}{\pa \bar{w}^{\gd_{1}}} \dots \frac{\pa }{\pa \bar{w}^{\gd_{j_{3}}}} \, . 
	\end{align}
\end{subequations}
The tensor $\cH$ is then extracted from the polynomial as follows:
\begin{equation}
	\cH_{\cA_{1} \cA_{2} \cA_{3}}(X) = \frac{\pa}{ \pa \boldsymbol{U}^{\cA_{1}} } \frac{\pa}{ \pa \boldsymbol{V}^{\cA_{2}}} \frac{\pa}{ \pa \boldsymbol{W}^{\cA_{3}} } \, \cH(X; U, V, W) \, .
\end{equation}
The polynomial $\cH$, \eqref{H - generating polynomial}, is now constructed out of scalar combinations of $X$, and the auxiliary spinors $U$, $V$ and $W$ with the appropriate homogeneity. Such a polynomial can be constructed out of the following monomials:
\begin{subequations} \label{Basis scalar structures}
	\begin{align} 
		P_{1} &= \ve_{\a \b} v^{\a} w^{\b} \, , & P_{2} &= \ve_{\a \b} w^{\a} u^{\b} \, , & P_{3} &= \ve_{\a \b} u^{\a} v^{\b} \, , \\
		\bar{P}_{1} &= \ve_{\ad \bd} \bar{v}^{\ad} \bar{w}^{\bd} \, , & \bar{P}_{2} &= \ve_{\ad \bd} \bar{w}^{\ad} \bar{u}^{\bd} \, , & \bar{P}_{3} &= \ve_{\ad \bd} \bar{u}^{\ad} \bar{v}^{\bd} \, , \\[1mm]
		Q_{1} &= \hat{X}_{\a \ad} \, v^{\a} \bar{w}^{\ad} \, ,  &  Q_{2} &= \hat{X}_{\a \ad} \, w^{\a} \bar{u}^{\ad} \, ,  &  Q_{3} &= \hat{X}_{\a \ad} \, u^{\a} \bar{v}^{\ad} \, , \\
		\bar{Q}_{1} &= \hat{X}_{\a \ad} \, w^{\a} \bar{v}^{\ad} \, ,  &  \bar{Q}_{2} &= \hat{X}_{\a \ad} \, u^{\a} \bar{w}^{\ad} \, ,  &  \bar{Q}_{3} &= \hat{X}_{\a \ad} \, v^{\a} \bar{u}^{\ad} \, , \\[1mm]
		Z_{1} &= \hat{X}_{\a \ad} \, u^{\a} \bar{u}^{\ad}  \, , & Z_{2} &= \hat{X}_{\a \ad} \, v^{\a} \bar{v}^{\ad} \, , & Z_{3} &= \hat{X}_{\a \ad} \, w^{\a} \bar{w}^{\ad} \, .
	\end{align}
\end{subequations}
To construct linearly independent structures for a given three-point function, one must also take into account the following linear dependence relations between the monomials:
\begin{subequations} \label{Linear dependence 1}
	\begin{align} 
		Z_{2} Z_{3} + P_{1} \bar{P}_{1} - Q_{1} \bar{Q}_{1} &= 0 \, , \\
		Z_{1} Z_{3} + P_{2} \bar{P}_{2} - Q_{2} \bar{Q}_{2} &= 0 \, , \\
		Z_{1} Z_{2} + P_{3} \bar{P}_{3} - Q_{3} \bar{Q}_{3} &= 0 \, ,
	\end{align}
\end{subequations}
\vspace{-10mm}
\begin{subequations} \label{Linear dependence 2}
	\begin{align} 
		Z_{1} P_{1} + P_{2} \bar{Q}_{3} + P_{3} Q_{2} &= 0 \, , & Z_{1} \bar{P}_{1} + \bar{P}_{2} Q_{3} + \bar{P}_{3} \bar{Q}_{2} &= 0  \, , \\
		Z_{2} P_{2} + P_{3} \bar{Q}_{1} + P_{1} Q_{3} &= 0 \, , & Z_{2} \bar{P}_{2} + \bar{P}_{3} Q_{1} + \bar{P}_{1} \bar{Q}_{3} &= 0 \, , \\
		Z_{3} P_{3} + P_{1} \bar{Q}_{2} + P_{2} Q_{1} &= 0 \, , & Z_{3} \bar{P}_{3} + \bar{P}_{1} Q_{2} + \bar{P}_{2} \bar{Q}_{1} &= 0 \, .
	\end{align}
\end{subequations}
\vspace{-10mm}
\begin{subequations} \label{Linear dependence 3}
	\begin{align} 
		Z_{1} Q_{1} + \bar{P}_{2} P_{3} - \bar{Q}_{2} \bar{Q}_{3} &= 0 \, , & Z_{1} \bar{Q}_{1} + P_{2} \bar{P}_{3} + Q_{2} Q_{3} &= 0  \, , \\
		Z_{2} Q_{2} + \bar{P}_{3} P_{1} - \bar{Q}_{3} \bar{Q}_{1} &= 0 \, , & Z_{2} \bar{Q}_{2} + P_{3} \bar{P}_{1} + Q_{3} Q_{1} &= 0 \, , \\
		Z_{3} Q_{3} + \bar{P}_{1} P_{2} - \bar{Q}_{1} \bar{Q}_{2} &= 0 \, , & Z_{3} \bar{Q}_{3} + P_{1} \bar{P}_{2} + Q_{1} Q_{2} &= 0 \, .
	\end{align}
\end{subequations}
These allow elimination of the combinations $Z_{i} Z_{j}$, $Z_{i} P_{i}$, $Z_{i} \bar{P}_{i}$, $Z_{i} Q_{i}$, $Z_{i} \bar{Q}_{i}$. There are also the following relations involving triple products:
\begin{subequations} \label{Linear dependence 4}
	\begin{align} 
		P_{1} \bar{P}_{2} \bar{P}_{3} + \bar{P}_{1} Q_{2} \bar{Q}_{3} + \bar{P}_{2} \bar{Q}_{3} \bar{Q}_{1} + \bar{P}_{3} Q_{1} Q_{2} &= 0 \, , \\
		P_{2} \bar{P}_{3} \bar{P}_{1} + \bar{P}_{2} Q_{3} \bar{Q}_{1} + \bar{P}_{3} \bar{Q}_{1} \bar{Q}_{2} + \bar{P}_{1} Q_{2} Q_{3} &= 0 \, , \\
		P_{3} \bar{P}_{1} \bar{P}_{2} + \bar{P}_{3} Q_{1} \bar{Q}_{2} + \bar{P}_{1} \bar{Q}_{2} \bar{Q}_{3} + \bar{P}_{2} Q_{3} Q_{1} &= 0 \, ,
	\end{align}
\end{subequations}
\vspace{-10mm}
\begin{subequations} \label{Linear dependence 5}
	\begin{align} 
		\bar{P}_{1} P_{2} P_{3} + P_{1} \bar{Q}_{2} Q_{3} + P_{2} Q_{3} Q_{1} + P_{3} \bar{Q}_{1} \bar{Q}_{2} &= 0 \, , \\
		\bar{P}_{2} P_{3} P_{1} + P_{2} \bar{Q}_{3} Q_{1} + P_{3} Q_{1} Q_{2} + P_{1} \bar{Q}_{2} \bar{Q}_{3} &= 0 \, , \\
		\bar{P}_{3} P_{1} P_{2} + P_{3} \bar{Q}_{1} Q_{2} + P_{1} Q_{2} Q_{3} + P_{2} \bar{Q}_{3} \bar{Q}_{1} &= 0 \, ,
	\end{align}
\end{subequations}
\vspace{-10mm}
\begin{subequations} \label{Linear dependence 6}
	\begin{align} 
		\bar{P}_{1} P_{2} \bar{Q}_{3} - P_{1} \bar{P}_{2} Q_{3} + \bar{Q}_{1} \bar{Q}_{2} \bar{Q}_{3} - Q_{1} Q_{2} Q_{3} &= 0 \, , \\
		\bar{P}_{2} P_{3} \bar{Q}_{1} - P_{2} \bar{P}_{3} Q_{1} + \bar{Q}_{1} \bar{Q}_{2} \bar{Q}_{3} - Q_{1} Q_{2} Q_{3} &= 0 \, , \\
		\bar{P}_{3} P_{1} \bar{Q}_{2} - P_{3} \bar{P}_{1} Q_{2} + \bar{Q}_{1} \bar{Q}_{2} \bar{Q}_{3} - Q_{1} Q_{2} Q_{3} &= 0 \, ,
	\end{align}
\end{subequations}
which allow for elimination of the products $P_{i} \bar{P}_{j} \bar{P}_{k}$, $\bar{P}_{i} P_{j} P_{k}$, $\bar{P}_{i} P_{j} \bar{Q}_{k}$. These relations (which appear to be exhaustive) are sufficient to reduce any set of structures in a given three-point function to a linearly independent set.

The task now is to construct a complete list of possible (linearly independent) solutions for the polynomial $\cH$ for a given set of spins. This process is simplified by introducing a generating function, $\cF(X; U, V, W \, | \, \G)$, defined as follows:
\begin{align} \label{Generating function}
		\cF(X; U,V,W \, | \, \G) &= P_{1}^{k_{1}} P_{2}^{k_{2}} P_{3}^{k_{3}} \, 	\bar{P}_{1}^{\bar{k}_{1}} \bar{P}_{2}^{\bar{k}_{2}} \bar{P}_{3}^{\bar{k}_{3}} Q_{1}^{l_{1}} Q_{2}^{l_{2}} Q_{3}^{l_{3}} \, \bar{Q}_{1}^{\bar{l}_{1}} \bar{Q}_{2}^{\bar{l}_{2}} \bar{Q}_{3}^{\bar{l}_{3}} Z_{1}^{r_{1}} Z_{2}^{r_{2}} Z_{3}^{r_{3}} \, ,
\end{align}
where the non-negative integers, $ \G =  \bigcup_{i \in \{1,2,3\} }  \{ k_{i}, \bar{k}_{i}, l_{i}, \bar{l}_{i}, r_{i}\}$, are solutions to the following linear system:
\begin{subequations} \label{Diophantine equations}
	\begin{align}
		k_{2} + k_{3} + r_{1} + l_{3} + \bar{l}_{2} &= i_{1} \, , &  \bar{k}_{2} + \bar{k}_{3} + r_{1} + \bar{l}_{3} + l_{2} &= j_{1} \, , \\
		k_{1} + k_{3} + r_{2} + l_{1} + \bar{l}_{3} &= i_{2} \, , &  \bar{k}_{1} + \bar{k}_{3} + r_{2} + \bar{l}_{1} + l_{3} &= j_{2} \, , \\
		k_{1} + k_{2} + r_{3} + l_{2} + \bar{l}_{1} &= i_{3} \, , &  \bar{k}_{1} + \bar{k}_{2} + r_{3} + \bar{l}_{2} + l_{1} &= j_{3} \, .
	\end{align}
\end{subequations}
Here $i_{1}, i_{2}, i_{3}$, $j_{1}, j_{2}, j_{3}$ are fixed integers corresponding to the spin representations of the fields in the three-point function. From here it is convenient to define
\begin{equation}
	\D s = \frac{1}{2}( i_{1} + i_{2} + i_{3} - j_{1} - j_{2} - j_{3} ) \, .
\end{equation}
Using the system of equations \eqref{Diophantine equations}, we obtain
\begin{align}
	\D s =  k_{1} + k_{2} + k_{3} - \bar{k}_{1} - \bar{k}_{2} - \bar{k}_{3}  \, ,
\end{align}
in addition to
\begin{subequations}
	\begin{align}
		&k_{1} + k_{2} + k_{3} \leq \min( i_{1} + i_{2}, i_{1} + i_{3}, i_{2} + i_{3} ) \, , \\
		&\bar{k}_{1} + \bar{k}_{2} + \bar{k}_{3} \leq \min( j_{1} + j_{2}, j_{1} + j_{3}, j_{2} + j_{3} ) \, .
	\end{align}
\end{subequations}
Hence, the conditions for a given three-point function to be non-vanishing are
\begin{align} \label{e1}
	- \min( j_{1} + j_{2}, j_{1} + j_{3}, j_{2} + j_{3} ) \leq \D s \leq \min( i_{1} + i_{2}, i_{1} + i_{3}, i_{2} + i_{3} ) \, .
\end{align}
Indeed, this is the same condition found in \cite{Elkhidir:2014woa}. Now given a finite number of solutions $\G_{I}$, $I = 1, ..., N$ to \eqref{Diophantine equations} for a particular choice of $i_{1}, i_{2}, i_{3}, j_{1}, j_{2}, j_{3}$, the most general ansatz for the polynomial $\cH$ in \eqref{H - generating polynomial} is as follows:
\begin{equation}
	\cH(X; U, V, W) = X^{\D_{3} - \D_{2} - \D_{1}} \sum_{I=1}^{N} a_{I} \cF(X; U, V, W \, | \, \G_{I}) \, ,
\end{equation}
where $a_{I}$ are a set of complex constants. Hence, constructing the most general ansatz for the generating polynomial $\cH$ is now equivalent to finding all non-negative integer solutions $\G_{I}$ of \eqref{Diophantine equations}. Once this ansatz has been obtained, the linearly independent structures can be found by systematically applying the linear dependence relations \eqref{Linear dependence 1}--\eqref{Linear dependence 6}.

Let us now recast the constraints on the three-point function into the auxiliary spinor formalism. Recalling that $s_{1} = \tfrac{1}{2}(i_{1} + j_{1})$, $s_{2} = \tfrac{1}{2}(i_{2} + j_{2})$, $s_{3} = \tfrac{1}{2}( i_{3} + j_{3})$, first we define:
\begin{subequations}
	\begin{align}
		J^{}_{s_{1}}(x_{1}; U) & = J^{}_{\a(i_{1}) \ad(j_{1})}(x_{1}) \, \boldsymbol{U}^{\a(i_{1}) \ad(j_{1})} \, , \\
		J'_{s_{2}}(x_{2}; V) &= J'_{\a(i_{2}) \ad(j_{2})}(x_{2}) \, \boldsymbol{V}^{\a(i_{2}) \ad(j_{2})} \, , \\
		J''_{s_{3}}(x_{3}; W) &= J''_{\g(i_{3}) \gd(j_{3})}(x_{3}) \, \boldsymbol{W}^{\g(i_{3}) \gd(j_{3})} \, ,
	\end{align}
\end{subequations}
where, to simplify notation, we denote $J^{}_{(s, q)} \equiv J^{}_{s}$. The general ansatz can be converted easily into the auxiliary spinor formalism, and is of the form
\begin{align}
	\begin{split}
		\langle J^{}_{s_{1}}(x_{1}; U) \, J'_{s_{2}}(x_{2}; V) \, J''_{s_{3}}(x_{3}; W) \rangle &= \frac{ \cI^{(i_{1}, j_{1})}(x_{13}; U, \tilde{U}) \,  \cI^{(i_{2}, j_{2})}(x_{23}; V, \tilde{V}) }{(x_{13}^{2})^{\D_{1}} (x_{23}^{2})^{\D_{2}}} \\
		& \hspace{5mm} \times \cH(X_{12}; \tilde{U},\tilde{V}, W) \, ,
	\end{split}
\end{align} 
where $\D_{i} = s_{i} + 2$. The generating polynomial, $\cH(X; U, V, W)$, is defined as
\begin{align}
	\cH(X; U,V,W) = \cH_{\a(i_{1}) \ad(j_{1}) \b(i_{2}) \bd(j_{2}) \g(i_{3}) \gd(j_{3})}(X) \, \boldsymbol{U}^{\a(i_{1}) \ad(j_{1})} \boldsymbol{V}^{\b(i_{2}) \bd(j_{2})} \boldsymbol{W}^{\g(i_{3}) \gd(j_{3})} \, ,
\end{align}
where
\begin{align}
	\cI^{(i,j)}(x; U, \tilde{U}) &\equiv \cI^{(i,j)}_{x}(U,\tilde{U}) = \boldsymbol{U}^{\a(i) \ad(j)} \cI_{\a(i)}{}^{\ad'(i)}(x) \, \bar{\cI}_{\ad(j)}{}^{\a'(j)}(x) \, \frac{\pa}{\pa \tilde{\boldsymbol{U}}^{\a'(j) \ad'(i)}} \, ,
\end{align}
is the inversion operator acting on polynomials degree $(i,j)$ in $(\tilde{\bar{u}}, \tilde{u})$. It should also be noted that $\tilde{\boldsymbol{U}}$ has index structure conjugate to $\boldsymbol{U}$. Sometimes we will omit the indices $(i,j)$ to streamline the notation. After converting the constraints summarised in the previous subsection into the auxiliary spinor formalism, we obtain:
\begin{enumerate}
	\item[\textbf{(i)}] {\bf Homogeneity:} \\
	Recall that $\cH$ is a homogeneous polynomial satisfying the following scaling property:
	\begin{align}
		\begin{split}
			\cH(\l^{2} X; U(i_{1},j_{1}), V(i_{2},j_{2}), W(i_{3},j_{3})) &= \\ 
			& \hspace{-35mm} (\l^{2})^{\D_{3} - \D_{2} - \D_{1}} \, \cH(X; U(i_{1},j_{1}), V(i_{2},j_{2}), W(i_{3},j_{3})) \, ,
		\end{split}
	\end{align}
	where we have used the notation $U(i_{1},j_{1})$, $V(i_{2},j_{2})$, $W(i_{3},j_{3})$ to keep track of homogeneity in the auxiliary spinors $(u, \bar{u})$, $(v,\bar{v})$ and $(w,\bar{w})$. For compactness we will suppress the homogeneities of the auxiliary spinors in the results.

	\item[\textbf{(ii)}] {\bf Differential constraints:} \\
	First, define the following three differential operators:
	\begin{align} \label{Derivative operators}
		D_{1} = \pa_{X}^{\a \ad} \frac{\pa}{\pa u^{\a}} \frac{\pa}{\pa \bar{u}^{\ad}} \, , && D_{2} =\pa_{X}^{\a \ad} \frac{\pa}{\pa v^{\a}} \frac{\pa}{\pa \bar{v}^{\ad}} \, , && D_{3} = \pa_{X}^{\a \ad} \frac{\pa}{\pa w^{\a}} \frac{\pa}{\pa \bar{w}^{\ad}} \, .
	\end{align}
	Conservation on all three points may be imposed using the following constraints:
	\begin{subequations} \label{Ch04.1-Conservation equations}
		\begin{align}
			\text{Conservation at $x_{1}$:} && D_{1} \, \cH(X; U,V,W) &= 0 \, , \\[1mm]
			\text{Conservation at $x_{2}$:} && D_{2} \, \cH(X; U,V,W) &= 0 \, , \\[1mm]
			\text{Conservation at $x_{3}$:} && D_{3} \, \tilde{\cH}(X; U,V,W) &= 0 \, ,
		\end{align}
	\end{subequations}
	where, in the auxiliary spinor formalism, $\tilde{\cH}$ is computed as follows:
	\begin{equation}
		\tilde{\cH}(X; U,V,W) = (X^{2})^{\D_{1} - \D_{3}} \cI_{X}(V,\tilde{V}) \, \cH^{c}(X; U, \tilde{V}, W) \, .
	\end{equation}
	Using the properties of the inversion tensor, it can be shown that this transformation is equivalent to the following replacement rules for the building blocks:
	\begin{subequations}
		\begin{align} 
			P_{1} &\rightarrow Q_{1} \, , & P_{2} &\rightarrow - \bar{P}_{2} \, , & P_{3} &\rightarrow - \bar{Q}_{3} \, \\
			\bar{P}_{1} &\rightarrow \bar{Q}_{1} \, , & \bar{P}_{2} &\rightarrow - P_{2} \, , & \bar{P}_{3} &\rightarrow - Q_{3} \, \\
			Q_{1} &\rightarrow - P_{1} \, , & Q_{2} &\rightarrow \bar{Q}_{2} \, , & Q_{3} &\rightarrow \bar{P}_{3} \, \\
			\bar{Q}_{1} &\rightarrow - \bar{P}_{1} \, , & \bar{Q}_{2} &\rightarrow Q_{2} \, , & \bar{Q}_{3} &\rightarrow P_{3} \, \\
			Z_{1} &\rightarrow Z_{1} \, , & Z_{2} &\rightarrow - Z_{2} \, , & Z_{3} &\rightarrow Z_{3} \, .
		\end{align}
	\end{subequations}

	\item[\textbf{(iii)}] {\bf Point switch symmetries:} \\
	If the fields $J$ and $J'$ coincide (hence $i_{1} = i_{2}$, $j_{1} = j_{2}$), then we obtain the following point-switch constraint
	\begin{equation} \label{Point switch A}
		\cH(X; U,V,W) = (-1)^{\e(J)} \cH(-X; V,U,W) \, ,
	\end{equation}
	where $\e(J)$ is the Grassmann parity of $J$. Similarly, if the fields $J$ and $J''$ coincide (hence $i_{1} = i_{3}$, $j_{1} = j_{3}$) then we obtain the constraint
	\begin{equation} \label{Point switch B}
		\tilde{\cH}(X; U, V, W) = (-1)^{\e(J)} \cH(- X; W, V, U) \, .
	\end{equation}

	\item[\textbf{(iv)}] {\bf Reality condition:} \\
	If the fields in the correlation function belong to the $(s,s)$ representation, then the three-point function must satisfy the reality condition
	\begin{equation} \label{Reality condition}
		\bar{\cH}(X; U, V, W) = \cH(X; U, V, W) \, .
	\end{equation}
	Similarly, if the fields at $J$, $J'$ at $x_{1}$ and $x_{2}$ respectively possess the same spin and are conjugate to each other, i.e. $J' = \bar{J}$, we must impose a combined reality/point-switch condition using the following constraint
	\begin{equation} \label{Reality/switch condition}
		\cH(X; U, V, W) = (-1)^{\epsilon(J)} \bar{\cH}(-X; V, U, W) \, ,
	\end{equation}
	where $\e(J)$ is the Grassmann parity of $J$.

\end{enumerate}
%


\subsubsection{Inversion transformation}\label{subsubsection2.2.3}


In general, whenever parity is a symmetry of a CFT, so too is invariance under inversions. 
An inversion transformation $\cI$ maps fields in the $(i, j)$ representation onto fields in the complex conjugate representation, $(j, i)$.\footnote{For a more detailed discussion of parity transformations in 4D CFT, see \cite{Elkhidir:2014woa}.} Hence, inversions map correlation functions of fields 
onto correlation functions of their complex conjugate fields. In particular, if the fields in a given three-point function belong to the $(s, s)$ representation then it is possible to 
construct linear combinations of structures for the three-point function which are eigenfunctions of the inversion operator. We denote these as parity-even and parity-odd solutions respectively.
Indeed, given a tensor $\cH_{\cA_{1} \cA_{2} \cA_{3}}(X) = X^{\D_{3} - \D_{2}- \D_{1}} \hat{\cH}_{\cA_{1} \cA_{2} \cA_{3}}(X)$, the following inversion formula holds:
\begin{align}
	\hat{\cH}^{c}_{\bar{\cA}_{1} \bar{\cA}_{2} \bar{\cA}_{3}}(X) = \bar{\cI}^{(1)}{}_{\bar{\cA}_{1}}{}^{\cA_{1}}(X) \, \bar{\cI}^{(2)}{}_{\bar{\cA}_{2}}{}^{\cA_{2}}(X) \, \bar{\cI}^{(3)}{}_{\bar{\cA}_{3}}{}^{\cA_{3}}(X) \, \hat{\cH}_{\cA_{1} \cA_{2} \cA_{3}}(-X) \, .
\end{align}
Hence, we notice that under $\cI$, $\cH$ transforms into the complex conjugate representation. An analogous formula can be derived using the auxiliary spinor formalism. Given a polynomial $\cH(X; U, V, W) = X^{\D_{3} - \D_{2}- \D_{1}} \hat{\cH}(X; U,V,W)$, the following holds:
\begin{align} \label{H inversion formula}
	\hat{\cH}^{c}(X; U, V, W) = \cI_{X}(U,\tilde{U}) \, \cI_{X}(V,\tilde{V}) \, \cI_{X}(W,\tilde{W}) \, \hat{\cH}(-X; \tilde{U}, \tilde{V}, \tilde{W}) \, .
\end{align}
It is easy to understand this formula as the monomials \eqref{Basis scalar structures} have simple transformation properties under $\cI$:
\begin{align} \label{Inversion of basis structures}
	P_{i} &\xrightarrow{\cI} - \bar{P}_{i} \, , & Q_{i} &\xrightarrow{\cI} \bar{Q}_{i} \, , & Z_{i} &\xrightarrow{\cI} Z_{i} \, ,
\end{align}
with analogous rules applying for the building blocks $\bar{P}_{i}, \bar{Q}_{i}$. Since, for primary fields in the $(s,s)$ representation the three-point maps onto itself under inversion, it is possible to classify the parity-even and parity-odd structures in $\cH$ using \eqref{H inversion formula}. By letting $\hat{\cH}(X) = \hat{\cH}^{(+)}(X) + \hat{\cH}^{(-)}(X)$, we have  
\begin{align} \label{H inversion classification}
	\hat{\cH}^{(\pm)}(X; U, V, W) = \pm \, \cI_{X}(U,\tilde{U}) \, \cI_{X}(V,\tilde{V}) \, \cI_{X}(W,\tilde{W}) \, \hat{\cH}^{(\pm)}(-X; \tilde{U}, \tilde{V}, \tilde{W}) \, .
\end{align}
Structures satisfying the above property are defined as parity-even/odd for overall sign $+/-$. This is essentially the same approach used to classify parity-even and parity-odd
three-point functions in 3D CFT \cite{Buchbinder:2022mys}, which proves to be equivalent to the classification based on the absence/presence of the Levi-Civita pseudo-tensor. 
However, it is crucial to note that in three dimensions the linearly independent basic monomial structures comprising $\cH$ are naturally eigenfunctions of the inversion operator. 
The same is not necessarily true for three-point functions in four dimensions due to \eqref{Inversion of basis structures}, as the monomials \eqref{Basis scalar structures} now map onto their complex conjugates. Hence, we are required to take non-trivial linear combinations of the basic structures
and use the linear dependence relations~\eqref{Linear dependence 1}--\eqref{Linear dependence 6} to form eigenfunctions of the inversion operator. Our classification of parity-even/odd solutions obtained this way is in complete agreement with the results found in~\cite{Stanev:2012nq}.


\section{Three-point functions of conserved currents}\label{section3}

In the next subsections we analyse the structure of three-point functions involving conserved currents in 4D CFT. 
We classify, using computational methods, all possible three-point functions involving the conserved currents $J_{(s,q)}$, $\bar{J}_{(s,q)}$ for $s_{i} \leq 10$. 
In particular, we determine the general structure and the number of independent solutions present in the three-point functions \eqref{possible three point functions}. 
As pointed out in the introduction, the number of independent conserved structures generically grows linearly with the minimum spin and the solution for the 
function $\cH(X; U, V, W)$ quickly becomes too long and complicated even for relatively low spins. 
Thus, although our method allows us to find $\cH(X; U, V, W)$ in a very explicit form for arbitrary spins (limited only by computer power), 
we find it practical to present the solutions when 
there is a small number of structures. Such examples involving low spins are discussed in subsection~\ref{subsection3.1}.
In subsection~\ref{subsection3.2} we state the classification for arbitrary spins. Some additional examples are presented in appendix \ref{AppB}.

\subsection{Conserved low-spin currents}\label{subsection3.1}

We begin our analysis by considering correlation functions involving conserved low-spin currents such as the energy-momentum tensor, vector current, and ``supersymmetry-like" currents in 4D CFT. Many of these results are known throughout the literature, but we derive them again here to demonstrate our approach.

\subsubsection{Energy-momentum tensor and vector current correlators}\label{subsubsection3.1.1}

The fundamental bosonic conserved currents in any conformal field theory are the conserved vector current, $V_{m}$, and the symmetric, traceless energy-momentum tensor, $T_{mn}$. The vector current has scale dimension $\Delta_{V} = 3$ and satisfies $\pa^{m} V_{m} = 0$, while the energy-momentum tensor has scale dimension $\Delta_{T} = 4$ and satisfies the conservation equation $\pa^{m} T_{mn} = 0$. Converting to spinor notation using the conventions outlined in appendix \ref{AppA}, we have:
\begin{align}
	V_{\a \ad}(x) = (\s^{m})_{\a \ad} V_{m}(x) \, , && T_{(\a_{1} \a_{2}) (\ad_{1} \ad_{2})}(x) = (\s^{m})_{(\a_{1} (\ad_{1}} (\g^{n})_{\a_{2}) \ad_{2})} T_{mn}(x) \, .
\end{align}
These objects possess fundamental information associated with internal and spacetime symmetries, hence, their three-point functions are of great importance. The possible three-point functions involving the conserved vector current and the energy-momentum tensor are:
\begin{subequations}
	\begin{align} \label{Low-spin component correlators}
		\langle V_{\a \ad}(x_{1}) \, V_{\b \bd}(x_{2}) \, V_{\g \gd}(x_{3}) \rangle \, , &&  \langle V_{\a \ad}(x_{1}) \, V_{\b \bd}(x_{2}) \, T_{\g(2) \gd(2)}(x_{3}) \rangle \, , \\
		\langle T_{\a(2) \ad(2)}(x_{1}) \, T_{\b(2) \bd(2)}(x_{2}) \, V_{\g \gd}(x_{3}) \rangle \, , &&  \langle T_{\a(2) \ad(2)}(x_{1}) \, T_{\b(2) \bd(2)}(x_{2}) \, T_{\g(2) \gd(2)}(x_{3}) \rangle \, .
	\end{align}
\end{subequations}
Let us first consider $\langle V V V \rangle$. By using the notation for the currents $J^{}_{(s,q)}$, $\bar{J}^{}_{(s,q)}$, this corresponds to the general three-point function $\langle J^{}_{(1,0)} J'_{(1,0)} J''_{(1,0)} \rangle$.\\


\noindent\textbf{Correlation function} $\langle J^{}_{(1,0)} J'_{(1,0)} J''_{(1,0)} \rangle$\textbf{:}\\[2mm]
The general ansatz for this correlation function, according to \eqref{Conserved correlator ansatz} is
\begin{align}
	\langle J^{}_{\a \ad}(x_{1}) \, J'_{\b \bd}(x_{2}) \, J''_{\g \gd}(x_{3}) \rangle = \frac{ \cI_{\a}{}^{\ad'}(x_{13}) \, \bar{\cI}_{\ad}{}^{\a'}(x_{13}) \,  \cI_{\b}{}^{\bd'}(x_{23}) \, \bar{\cI}_{\bd}{}^{\b'}(x_{23}) }{(x_{13}^{2})^{3} (x_{23}^{2})^{3}}
	\; \cH_{\a' \ad' \b' \bd' \g \gd}(X_{12}) \, .
\end{align} 
Using the formalism outlined in subsection \ref{subsubsection2.2.2}, all information about this correlation function is encoded in the following polynomial:
\begin{align}
	\cH(X; U, V, W) = \cH_{ \a \ad \b \bd \g \gd }(X) \, \boldsymbol{U}^{\a \ad}  \boldsymbol{V}^{\b \bd}  \boldsymbol{W}^{\g \gd} \, .
\end{align}
Using Mathematica we solve \eqref{Diophantine equations} for the chosen spin representations of the currents and substitute each solution into the generating function \eqref{Generating function}. This provides us with the following list of (linearly dependent) polynomial structures:
%
\begin{align}
	\begin{split}
		&\big\{ Q_1 Q_2 Q_3,Z_1 Z_2 Z_3,P_3 Q_2 \bar{P}_1,P_1 Z_1 \bar{P}_1,P_1 Q_3 \bar{P}_2,P_2 Z_2 \bar{P}_2,P_2 Q_1 \bar{P}_3, P_3 Z_3 \bar{P}_3, \\ 
		& \hspace{20mm}  Q_1 Z_1 \bar{Q}_1,P_3 \bar{P}_2 \bar{Q}_1, Q_2 Z_2 \bar{Q}_2,P_1 \bar{P}_3 \bar{Q}_2,Q_3 Z_3 \bar{Q}_3,P_2 \bar{P}_1
		\bar{Q}_3,\bar{Q}_1 \bar{Q}_2 \bar{Q}_3 \big\}\,.
	\end{split}
\end{align}
Next, we systematically apply the linear dependence relations \eqref{Linear dependence 1} to these lists, reducing them to the following sets of linearly independent structures:
%
\begin{align}
	\big\{ Q_1 Q_2 Q_3,P_3 Q_2 \bar{P}_1,P_1 Q_3 \bar{P}_2,P_2 Q_1 \bar{P}_3,\bar{Q}_1 \bar{Q}_2 \bar{Q}_3 \big\}\,.
\end{align}
Note that application of the linear-dependence relations eliminates all terms involving $Z_{i}$ in this case.  Since this correlation function is composed of fields in the $(s,s)$ representation, the solutions for the three-point function may be split up into parity-even and parity-odd contributions. To do this we construct linear combinations for the polynomial $\hat{\cH}(X; U, V, W)$ which are even/odd under inversion in accordance with \eqref{H inversion classification}:
%
%
\begin{align}
	\begin{split}
		&A_1  (\bar{Q}_1 \bar{Q}_2 \bar{Q}_3+Q_1
		Q_2 Q_3 ) + A_2  (P_3 Q_2 \bar{P}_1-\bar{Q}_1 \bar{Q}_2 \bar{Q}_3 ) +A_3  (P_1 Q_3 \bar{P}_2-\bar{Q}_1 \bar{Q}_2
		\bar{Q}_3 ) \\
		& \hspace{20mm} +A_4  (P_2 Q_1 \bar{P}_3-\bar{Q}_1 \bar{Q}_2 \bar{Q}_3 )+ B_1  (Q_1 Q_2 Q_3-\bar{Q}_1 \bar{Q}_2 \bar{Q}_3 ) \, .
	\end{split}
\end{align}
We note here (and in all other examples) that the parity-even contributions possess the complex coefficients $A_{i}$, while the parity-odd solutions possess the complex coefficients $B_{i}$. It can be explicitly checked that these structures possess the appropriate transformation properties. Next, since the correlation function is overall real, we must impose the reality condition \eqref{Reality condition}. As a result, we find that the parity-even coefficients $A_{i}$ are purely real, i.e., $A_{i} = a_{i}$, while the parity-odd coefficients $B_{i}$ are purely imaginary, i.e., $B_{i} = \text{i} b_{i}$.

We must now impose the conservation of the currents. Following the procedure outlined in \ref{subsubsection2.2.2} we obtain a linear system in the coefficients $a_{i}$, $b_{i}$ which can be easily solved computationally. We find the following solution for $\cH(X;U,V,W)$ consistent with conservation on all three points:
%
%
\begin{align}
	\begin{split}
		& \frac{a_1}{X^3} \big(Q_1 Q_2 Q_3 + 2 P_1 Q_3 \bar{P}_2-\bar{Q}_1 \bar{Q}_2 \bar{Q}_3 \big) \\[1mm]
		& \hspace{10mm} +\frac{a_2}{X^3} \big( P_3 Q_2
		\bar{P}_1-3 P_1 Q_3 \bar{P}_2+P_2 Q_1 \bar{P}_3+\bar{Q}_1 \bar{Q}_2 \bar{Q}_3 \big) \\[1mm]
		& \hspace{25mm} + \frac{\text{i} b_1}{X^3} \big( Q_1 Q_2 Q_3- \bar{Q}_1 \bar{Q}_2
		\bar{Q}_3 \big) \, .
	\end{split}
\end{align}
The only remaining constraints to impose are symmetries under permutations of spacetime points, which apply when the currents in the three-point function are identical, i.e. when $J=J'=J''$. After imposing \eqref{Point switch A}, \eqref{Point switch B}, only the structure corresponding to the coefficient $b_{1}$ survives. 
However, the $a_{1}$, $a_{2}$ structures can exist if the currents are non-abelian. 
This is consistent with the results of \cite{Osborn:1993cr, Erdmenger:1996yc, Stanev:2012nq}.\footnote{The coefficient $b_1$ is related to the chiral anomaly of the CFT under consideration when it is coupled to a background vector field \cite{Erdmenger:1996yc}. This anomaly exists in chiral theories which are not invariant under parity and, thus, admit a parity-odd contribution.}

The next example to consider is the mixed correlator $\langle V V T \rangle$. To study this case we may examine the correlation function 
$\langle J^{}_{(1,0)} J'_{(1,0)} J''_{(2,0)} \rangle$. \\[5mm]
\noindent
\textbf{Correlation function} $\langle J^{}_{(1,0)} J'_{(1,0)} J''_{(2,0)} \rangle$\textbf{:}\\[2mm]
Using the general formula, the ansatz for this three-point function is:
\begin{align}
	\begin{split}
		\langle J^{}_{\a \ad}(x_{1}) \, J'_{\b \bd}(x_{2}) \, J''_{\g(2) \gd(2)}(x_{3}) \rangle &= \frac{ \cI_{\a}{}^{\ad'}(x_{13}) \, \bar{\cI}_{\ad}{}^{\a'}(x_{13}) \,  \cI_{\b}{}^{\bd'}(x_{23}) \, \bar{\cI}_{\bd}{}^{\b'}(x_{23})}{(x_{13}^{2})^{3} (x_{23}^{2})^{3}} \\
		& \hspace{20mm} \times \, \cH_{\a' \ad' \b' \bd' \g(2) \gd(2)}(X_{12}) \, .
	\end{split}
\end{align} 
All information about this correlation function is encoded in the following polynomial:
\begin{align}
	\cH(X; U, V, W) = \cH_{ \a \ad \b \bd \g(2) \gd(2) }(X) \, \boldsymbol{U}^{\a \ad}  \boldsymbol{V}^{\b \bd}  \boldsymbol{W}^{\g(2) \gd(2)} \, .
\end{align}
After solving \eqref{Diophantine equations}, we find the following linearly dependent polynomial structures:
%
\begin{align}
	\begin{split}
		&\big\{ P_1 P_2 \bar{P}_1 \bar{P}_2,P_2 Q_1 Q_2 \bar{P}_1,P_3 Q_2 Z_3 \bar{P}_1,P_3 Z_3^2 \bar{P}_3,  Q_3 Z_3^2 \bar{Q}_3, Z_1 Z_2 Z_3^2,\\
		& \hspace{10mm} P_1 Q_3 Z_3 \bar{P}_2, P_2 Z_2 Z_3 \bar{P}_2, Q_1 Q_2 Q_3 Z_3, P_2 Q_1 Z_3 \bar{P}_3,Q_1 Z_1 Z_3 \bar{Q}_1, \\
		& \hspace{20mm} P_2 Q_1 \bar{P}_2 \bar{Q}_1, P_3 Z_3 \bar{P}_2 \bar{Q}_1,Q_2 Z_2 Z_3 \bar{Q}_2,P_1 Q_2 \bar{P}_1 \bar{Q}_2,P_1 Z_3 \bar{P}_3
		\bar{Q}_2, \\
		& \hspace{30mm} Q_1 Q_2 \bar{Q}_1 \bar{Q}_2, P_1 \bar{P}_2 \bar{Q}_1 \bar{Q}_2,P_1 Z_1 Z_3 \bar{P}_1,P_2 Z_3 \bar{P}_1 \bar{Q}_3,Z_3
		\bar{Q}_1 \bar{Q}_2 \bar{Q}_3 \big\}\,.
\end{split}
\end{align}
We now systematically apply the linear dependence relations \eqref{Linear dependence 1}--\eqref{Linear dependence 6} to obtain the linearly independent structures
%
\begin{align}
	\big\{ P_2 Q_1 Q_2 \bar{P}_1,P_1 P_2 \bar{P}_1 \bar{P}_2,P_2 Q_1 \bar{P}_2 \bar{Q}_1,P_1 Q_2 \bar{P}_1 \bar{Q}_2,Q_1 Q_2
	\bar{Q}_1 \bar{Q}_2,P_1 \bar{P}_2 \bar{Q}_1 \bar{Q}_2 \big\}\,.
\end{align}
Next, we construct the following parity-even and parity-odd linear combinations which comprise the polynomial $\hat{\cH}(X; U, V, W)$:
%
\begin{align}
	\begin{split}
		& A_1  (P_2 Q_1 Q_2 \bar{P}_1+P_1 \bar{P}_2 \bar{Q}_1
		\bar{Q}_2 )+A_2 P_1 P_2 \bar{P}_1 \bar{P}_2+A_3 P_2 Q_1 \bar{P}_2 \bar{Q}_1 \\
		& \hspace{5mm} +A_4 P_1 Q_2 \bar{P}_1 \bar{Q}_2+A_5 Q_1 Q_2 \bar{Q}_1 \bar{Q}_2+B_1 (P_2 Q_1 Q_2 \bar{P}_1-P_1 \bar{P}_2
		\bar{Q}_1 \bar{Q}_2 ) \, .
	\end{split}
\end{align}
%
We now impose conservation on all three points to obtain the final solution for $\cH(X;U,V,W)$
\begin{align}
	\begin{split}
		&\frac{a_1}{X^2} \Big( P_2 Q_1 Q_2 \bar{P}_1+P_1 Q_2 \bar{P}_1
		\bar{Q}_2+P_2 Q_1 \bar{P}_2 \bar{Q}_1+P_1 \bar{P}_2 \bar{Q}_1 \bar{Q}_2-\sfrac{2}{3} Q_1 Q_2 \bar{Q}_1
		\bar{Q}_2 \Big) \\[1mm]
		& \hspace{5mm} +\frac{a_2}{X^2} \Big(-\sfrac{1}{2} P_2 Q_1 \bar{P}_2 \bar{Q}_1-\sfrac{1}{2} P_1 Q_2 \bar{P}_1 \bar{Q}_2+P_1 P_2 \bar{P}_1
		\bar{P}_2+\sfrac{1}{3} Q_1 Q_2 \bar{Q}_1 \bar{Q}_2 \Big) \\[1mm]
		& \hspace{15mm} +\frac{\text{i} b_1}{X^2} \Big( P_2 Q_1 Q_2 \bar{P}_1- P_1 \bar{P}_2 \bar{Q}_1 \bar{Q}_2 \Big) \, .
	\end{split}
\end{align}
In this case, only the parity-even structures (proportional to $a_{1}$ and $a_{2}$) survive after setting $J = J'$. Hence, this correlation function is fixed up to two independent 
parity-even structures with real coefficients.

The number of polynomial structures increases rapidly for increasing $s_{i}$, and for the three-point functions $\langle T T V \rangle$, $\langle T T T \rangle$ we will present only the linearly independent structures and the final results after imposing parity, reality, and conservation on all three points. For $\langle T T V \rangle$ we may consider the correlation function $\langle J^{}_{(2,0)} J'_{(2,0)} J''_{(1,0)} \rangle$, which is constructed from the following list of linearly independent structures:
\begin{align}
	\begin{split}
		& \big\{ P_3 Q_1 Q_2 Q_3 \bar{P}_3,P_3^2 Q_2 \bar{P}_1 \bar{P}_3,P_2 P_3 Q_1 \bar{P}_3^2,Q_1 Q_2 Q_3^2 \bar{Q}_3, P_1 Q_3^2 \bar{P}_2 \bar{Q}_3, \\
		& \hspace{20mm} P_3 Q_2 Q_3 \bar{P}_1 \bar{Q}_3, P_2 Q_1 Q_3 \bar{P}_3 \bar{Q}_3,P_3 \bar{P}_3 \bar{Q}_1 \bar{Q}_2
		\bar{Q}_3,Q_3 \bar{Q}_1 \bar{Q}_2 \bar{Q}_3^2 \big\}\,.
	\end{split}
\end{align}
We now construct linearly independent parity-even and parity-odd solutions consistent with \eqref{H inversion classification}. Then, after imposing all the constraints due to reality and conservation, we obtain the final solution for $\cH(X; U,V,W)$:
\begin{align}
	\begin{split}
		&\frac{a_1}{X^5} \Big(3 P_1 Q_3^2 \bar{P}_2 \bar{Q}_3+P_3 Q_1 Q_2 Q_3 \bar{P}_3+2 P_3 Q_2 Q_3 \bar{P}_1 \bar{Q}_3 \\
		& \hspace{10mm} +2 P_2 Q_1 Q_3
		\bar{P}_3 \bar{Q}_3 +P_3 \bar{P}_3 \bar{Q}_1 \bar{Q}_2 \bar{Q}_3+\sfrac{7}{2} Q_1 Q_2 Q_3^2 \bar{Q}_3-\sfrac{7}{2} Q_3 \bar{Q}_1
		\bar{Q}_2 \bar{Q}_3^2 \Big) \\[1mm]
		&+\frac{a_2}{X^5} \Big(P_3^2 Q_2 \bar{P}_1 \bar{P}_3+P_2 P_3 Q_1 \bar{P}_3^2-6 P_3 Q_2 Q_3
		\bar{P}_1 \bar{Q}_3-2 P_3 \bar{P}_3 \bar{Q}_1 \bar{Q}_2 \bar{Q}_3 \\
		& \hspace{10mm}-7 P_1 Q_3^2 \bar{P}_2 \bar{Q}_3-6 P_2 Q_1 Q_3 \bar{P}_3
		\bar{Q}_3+\sfrac{17}{2} Q_3 \bar{Q}_1 \bar{Q}_2 \bar{Q}_3^2-\sfrac{21}{2} Q_1 Q_2 Q_3^2 \bar{Q}_3 \Big) \\[1mm]
		& \hspace{5mm} +\frac{\text{i} b_1}{X^5} \Big( 
		P_3 Q_1 Q_2 Q_3 \bar{P}_3- P_3 \bar{P}_3 \bar{Q}_1 \bar{Q}_2 \bar{Q}_3-\sfrac{3}{2} Q_1 Q_2 Q_3^2 \bar{Q}_3+\sfrac{3}{2} 
		Q_3 \bar{Q}_1 \bar{Q}_2 \bar{Q}_3^2 \Big) \, .
	\end{split}
\end{align}
After setting $J = J'$ and imposing the required symmetries under the exchange of $x_{1}$ and $x_{2}$ we find that $b_{1} = 0$, while $a_{1}, a_{2}$ remain unconstrained. Hence, the correlation function $\langle T T V \rangle$ is fixed up to two parity-even structures with real coefficients. 

The final fundamental three-point function to study is $\langle T T T \rangle$, and for this we analyse the correlation function $\langle J^{}_{(2,0)} J'_{(2,0)} J''_{(2,0)} \rangle$. In this case there are 15 linearly independent structures to consider:
\begin{align}
	\begin{split}
		&\big\{Q_1^2 Q_2^2 Q_3^2,P_3 Q_1 Q_2^2 Q_3 \bar{P}_1,P_3^2 Q_2^2 \bar{P}_1^2,P_1 Q_1 Q_2 Q_3^2 \bar{P}_2,P_1^2 Q_3^2
		\bar{P}_2^2,P_2 Q_1^2 Q_2 Q_3 \bar{P}_3, \\
		& \hspace{10mm} P_3 Q_1 Q_2 \bar{P}_3 \bar{Q}_1 \bar{Q}_2,P_2 Q_1 Q_3
		\bar{P}_2 \bar{Q}_1 \bar{Q}_3,P_1 Q_2 Q_3 \bar{P}_1 \bar{Q}_2 \bar{Q}_3,Q_1 Q_2 Q_3 \bar{Q}_1 \bar{Q}_2 \bar{Q}_3, \\
		& \hspace{15mm} P_2^2 Q_1^2 \bar{P}_3^2,P_3 Q_2 \bar{P}_1 \bar{Q}_1 \bar{Q}_2 \bar{Q}_3,P_1 Q_3 \bar{P}_2 \bar{Q}_1 \bar{Q}_2 \bar{Q}_3,P_2 Q_1 \bar{P}_3 \bar{Q}_1 \bar{Q}_2
		\bar{Q}_3,\bar{Q}_1^2 \bar{Q}_2^2 \bar{Q}_3^2 \big\}\,.
	\end{split}
\end{align}
From these structures we construct linear combinations that are even/odd under parity, analogous to the previous examples. Then, after imposing reality and conservation on all three points we obtain the following solution for $\cH(X; U,V,W)$:
%
\begin{align}
	\begin{split}
		&\frac{a_1}{X^4}
		\Big(Q_1^2 Q_2^2 Q_3^2+2 P_1^2 Q_3^2 \bar{P}_2^2 -2 Q_1 Q_2  Q_3 \bar{Q}_1 \bar{Q}_2 \bar{Q}_3 \\
		& \hspace{10mm}+2 P_1 Q_1 Q_2  Q_3^2 \bar{P}_2 -2
		P_1 Q_3 \bar{P}_2 \bar{Q}_1 \bar{Q}_2 \bar{Q}_3 +\bar{Q}_1^2 \bar{Q}_2^2 \bar{Q}_3^2 \Big) \\[1mm]
		&+\frac{a_2}{X^4} \Big(P_2 Q_1^2 Q_2 Q_3 \bar{P}_3 +P_3 Q_1 Q_2^2 Q_3 \bar{P}_1 -\sfrac{17}{3} P_1 Q_1 Q_2 Q_3^2 \bar{P}_2 +2 P_2 Q_1 Q_3
		\bar{P}_2 \bar{Q}_1 \bar{Q}_3  \\
		& \hspace{10mm} +3 Q_1 Q_2 Q_3 \bar{Q}_1 \bar{Q}_2 \bar{Q}_3+P_2 Q_1 \bar{P}_3 \bar{Q}_1 \bar{Q}_2 \bar{Q}_3
		-\sfrac{20}{3} P_1^2 Q_3^2 \bar{P}_2^2-3 \bar{Q}_1^2 \bar{Q}_2^2 \bar{Q}_3^2 \\
		& \hspace{15mm} +2 P_1 Q_2 Q_3 \bar{P}_1 \bar{Q}_2 \bar{Q}_3+P_3
		Q_2 \bar{P}_1 \bar{Q}_1 \bar{Q}_2 \bar{Q}_3+\sfrac{23}{3} P_1 Q_3 \bar{P}_2 \bar{Q}_1 \bar{Q}_2
		\bar{Q}_3 \Big) \\[1mm]
		&+\frac{a_3}{X^4} \Big( P_3^2 Q_2^2 \bar{P}_1^2 + \sfrac{19}{3}
		P_1^2 Q_3^2 \bar{P}_2^2  +\sfrac{16}{3} P_1 Q_1 Q_2 Q_3^2 \bar{P}_2  -2 Q_1 Q_2 Q_3 \bar{Q}_1 \bar{Q}_2 \bar{Q}_3  \\
		& \hspace{8mm} + P_2^2 Q_1^2 \bar{P}_3^2 -2 P_2 Q_1 Q_3 \bar{P}_2 \bar{Q}_1 \bar{Q}_3  -2 P_1 Q_2 Q_3 \bar{P}_1 \bar{Q}_2 \bar{Q}_3  -\sfrac{22}{3} P_1 Q_3 \bar{P}_2 \bar{Q}_1 \bar{Q}_2
		\bar{Q}_3  \\
		& \hspace{13mm} -3 P_3 Q_1 Q_2 \bar{P}_3
		\bar{Q}_1 \bar{Q}_2-2 P_3 Q_2 \bar{P}_1 \bar{Q}_1 \bar{Q}_2 \bar{Q}_3-2 P_2 Q_1 \bar{P}_3 \bar{Q}_1 \bar{Q}_2 \bar{Q}_3 +3 \bar{Q}_1^2 \bar{Q}_2^2 \bar{Q}_3^2 \Big) \\[1mm]
		&+\frac{\text{i} b_1}{X^4} \Big( Q_1^2 Q_2^2 Q_3^2+2  P_1 Q_1 Q_2 Q_3^2 \bar{P}_2 + \bar{Q}_1^2 \bar{Q}_2^2 \bar{Q}_3^2 \\
		& \hspace{25mm} -2  Q_1 Q_2 Q_3 \bar{Q}_1 \bar{Q}_2 \bar{Q}_3 -2  P_1 Q_3 \bar{P}_2 \bar{Q}_1 \bar{Q}_2 \bar{Q}_3 \Big) \\[1mm]
		&+\frac{\text{i} b_2}{X^4} \Big( P_2 Q_1^2 Q_2 Q_3 \bar{P}_3 + P_3 Q_1 Q_2^2 Q_3 \bar{P}_1 - 3 P_1 Q_1 Q_2 Q_3^2 \bar{P}_2 + Q_1 Q_2 Q_3 \bar{Q}_1 \bar{Q}_2 \bar{Q}_3 \\
		& \hspace{15mm} - P_2 Q_1 \bar{P}_3 \bar{Q}_1 \bar{Q}_2 \bar{Q}_3 - P_3 Q_2 \bar{P}_1 \bar{Q}_1 \bar{Q}_2 \bar{Q}_3+3  P_1 Q_3 \bar{P}_2 \bar{Q}_1 \bar{Q}_2 \bar{Q}_3 - \bar{Q}_1^2 \bar{Q}_2^2
		\bar{Q}_3^2 \Big) \, .
	\end{split}
\end{align}
In this case only three of the structures (corresponding to the real coefficients $a_{1}, a_{2}, a_{3}$) survive the point-switch symmetries upon exchange of $x_{1}$, $x_{2}$ and $x_{3}$. Hence, $\langle T T T \rangle$ is fixed up to three parity-even structures with real coefficients.

In all cases we note that the number of independent structures (prior to imposing exchange symmetries) is $2 \min(s_{1}, s_{2}, s_{3}) + 1$ in general, 
where $\min(s_{1}, s_{2}, s_{3}) + 1$ are parity-even and $\min(s_{1}, s_{2}, s_{3})$ are parity-odd.
 These results are in agreement with \cite{Osborn:1993cr,Stanev:2012nq, Stanev:2013eha, Stanev:2013qra, Zhiboedov:2012bm} in terms of the number of independent 
 structures, however, our construction of the three-point function is quite different.

\subsubsection{Spin-3/2 current correlators}\label{subsubsection3.1.2}

In this section we will evaluate three-point functions involving conserved fermionic currents. The most important examples of fermionic conserved currents in 4D CFT are the supersymmetry currents, $Q_{m,\a}$, $\bar{Q}_{m,\ad}$, which appear in $\cN$-extended superconformal field theories. Such fields are primary with dimension $\D_{Q} = \D_{\bar{Q}} = 7/2$, and satisfy the conservation equations $\pa^{m} Q_{m, \a } = 0$, $\pa^{m} \bar{Q}_{m, \ad } = 0$. In spinor notation, we have:
\begin{equation}
	Q_{\a \ad, \b}(x) = (\s^{m})_{\a \ad} Q_{m,\b}(x) \, , \hspace{10mm} \bar{Q}_{\a \ad, \bd}(x) = (\s^{m})_{\a \ad} \bar{Q}_{m,\bd}(x) \, .
\end{equation}
%
The correlation functions involving supersymmetry currents, vector currents, and the energy-momentum tensor are of fundamental importance. 
The four possible three-point functions involving $Q$, $V$ and $T$ which are of interest in $\cN=1$ superconformal field theories are:
\begin{subequations} \label{Susy current correlators - 1}
	\begin{align}
		\langle Q_{\a(2) \ad}(x_{1}) \, Q_{\b(2) \bd}(x_{2}) \, V_{\g \gd}(x_{3}) \rangle \, , && \langle Q_{\a(2) \ad}(x_{1}) \,  Q_{\b(2) \bd}(x_{2}) \, T_{\g(2) \gd(2)}(x_{3}) \rangle \, , \\
		\langle Q_{\a(2) \ad}(x_{1}) \, \bar{Q}_{\b \bd(2)}(x_{2}) \, V_{\g \gd}(x_{3}) \rangle \, , && \langle Q_{\a(2) \ad}(x_{1}) \,  \bar{Q}_{\b(2) \bd}(x_{2}) \, T_{\g(2) \gd(2)}(x_{3}) \rangle \, .
	\end{align}
\end{subequations}
%

These three-point functions were analysed in \cite{Buchbinder:2022cqp} using a similar approach, but we present them again here for completeness and to demonstrate our general formalism. Note that in the subsequent analysis we assume only conformal symmetry, not supersymmetry. 

We now present an explicit analysis of the general structure of correlation functions involving $Q$, $\bar{Q}$, $V$ and $T$ that are compatible with the constraints of conformal symmetry and conservation equations. Using our conventions for the currents, we recall that $Q \equiv J_{(3/2,1)}$, $\bar{Q} \equiv \bar{J}_{(3/2,1)}$. Let us first consider $\langle Q Q V \rangle$, for which we may analyse the general structure of the correlation function $\langle J^{}_{(3/2,1)} J'_{(3/2,1)} J''_{(1,0)} \rangle$. \\[5mm]
\noindent
\textbf{Correlation function} $\langle J^{}_{(3/2,1)} J'_{(3/2,1)} J''_{(1,0)} \rangle$\textbf{:}\\[2mm]
Using the general formula, the ansatz for this three-point function:
\begin{align}
	\begin{split}
		\langle J^{}_{\a(2) \ad}(x_{1}) \, J'_{\b(2) \bd}(x_{2}) \, J''_{\g \gd}(x_{3}) \rangle &= \frac{ \cI_{\a(2)}{}^{\ad'(2)}(x_{13}) \, \bar{\cI}_{\ad}{}^{\a'}(x_{13} ) \,  \cI_{\b(2)}{}^{\bd'(2)}(x_{23}) \,  \bar{\cI}_{\bd}{}^{\b'}(x_{23}) }{(x_{13}^{2})^{7/2} (x_{23}^{2})^{7/2}} \\ 
		& \hspace{5mm} \times \cH_{\a' \ad'(2) \b' \bd'(2) \g \gd}(X_{12}) \, .
	\end{split}
\end{align} 
Using the formalism outlined in \ref{subsubsection2.2.2}, all information about this correlation function is encoded in the following polynomial:
\begin{align}
	\cH(X; U, V, W) = \cH_{ \a \ad(2) \b \bd(2) \g \gd }(X) \, \boldsymbol{U}^{\a \ad(2)}  \boldsymbol{V}^{\b \bd(2)}  \boldsymbol{W}^{\g \gd} \, .
\end{align}
After solving \eqref{Diophantine equations}, we find the following linearly dependent polynomial structures in the even and odd sectors respectively:
%
\begin{align}
	\begin{split}
		&\big\{Q_2 Z_1 Z_2 \bar{P}_1,Q_2 Q_3 Z_2 \bar{P}_2,Q_1 Q_2 Q_3 \bar{P}_3,Z_1 Z_2 Z_3 \bar{P}_3,P_3 Q_2 \bar{P}_1 \bar{P}_3, \\
		& \hspace{5mm} P_1 Z_1 \bar{P}_1 \bar{P}_3,P_1 Q_3 \bar{P}_2 \bar{P}_3, P_2 Z_2 \bar{P}_2 \bar{P}_3,P_2 Q_1 \bar{P}_3^2,P_3 Z_3 \bar{P}_3^2, P_1 \bar{P}_3^2 \bar{Q}_2, \\
		& \hspace{10mm} Q_1 Z_1 \bar{P}_3 \bar{Q}_1,P_3 \bar{P}_2 \bar{P}_3 \bar{Q}_1,Q_2 Z_2 \bar{P}_3 \bar{Q}_2, Z_1 Z_2 \bar{P}_2 \bar{Q}_1, Q_2 Q_3 \bar{P}_1 \bar{Q}_3,\\
		& \hspace{15mm} Q_3 Z_3 \bar{P}_3 \bar{Q}_3,P_2 \bar{P}_1 \bar{P}_3 \bar{Q}_3,Z_1 \bar{P}_1
		\bar{Q}_1 \bar{Q}_3,Q_3 \bar{P}_2 \bar{Q}_1 \bar{Q}_3,\bar{P}_3 \bar{Q}_1 \bar{Q}_2 \bar{Q}_3 \big\}\,.
	\end{split}
\end{align}
Next we systematically apply the linear dependence relations \eqref{Linear dependence 1}--\eqref{Linear dependence 6} and obtain the following linearly independent structures:
%
\begin{align}
	\big\{ Q_1 Q_2 Q_3 \bar{P}_3,P_3 Q_2 \bar{P}_1 \bar{P}_3,P_2 Q_1 \bar{P}_3^2,Q_2 Q_3 \bar{P}_1 \bar{Q}_3,Q_3 \bar{P}_2 \bar{Q}_1
	\bar{Q}_3,\bar{P}_3 \bar{Q}_1 \bar{Q}_2 \bar{Q}_3 \big\}\,.
\end{align}
We now impose conservation on all three points and find that the solution for $\cH(X; U,V,W)$ is unique up to a complex coefficient, $A_{1} = a_{1} + \text{i} \tilde{a}_{1}$:
%
\begin{align} \label{QQV}
	\begin{split}
		&\frac{A_1}{X^4} \Big(Q_1 Q_2 Q_3 \bar{P}_3 + \sfrac{5}{9} P_2 Q_1 \bar{P}_3^2+\sfrac{5}{9} P_3 Q_2 \bar{P}_1 \bar{P}_3 \\
		& \hspace{15mm} -\sfrac{1}{9} \bar{P}_3 \bar{Q}_1 \bar{Q}_2 \bar{Q}_3-\sfrac{2}{9} Q_2 Q_3 \bar{P}_1 \bar{Q}_3-\sfrac{2}{9} Q_3 \bar{P}_2 \bar{Q}_1
		\bar{Q}_3 \Big) \, .
	\end{split}
\end{align}
However, this three-point function is not compatible with the point-switch symmetry associated with setting $J=J'$. Therefore we conclude that the three-point function $\langle Q Q V \rangle$ must vanish in general. 

\vspace{2mm}

\noindent
\textbf{Correlation function} $\langle J^{}_{(3/2,1)} \bar{J}'_{(3/2,1)} J''_{(1,0)} \rangle$\textbf{:}\\[2mm]
Using the general formula we obtain the following ansatz:
\begin{align}
	\begin{split}
		\langle J^{}_{\a(2) \ad}(x_{1}) \, J'_{\b \bd(2)}(x_{2}) \, J''_{\g \gd}(x_{3}) \rangle &= \frac{ \cI_{\a(2)}{}^{\ad'(2)}(x_{13}) \, \bar{\cI}_{\ad}{}^{\a'}(x_{13}) \,  \cI_{\b}{}^{\bd'}(x_{23}) \,  \bar{\cI}_{\bd(2)}{}^{\b'(2)}(x_{23}) }{(x_{13}^{2})^{7/2} (x_{23}^{2})^{7/2}} \\
		& \hspace{15mm} \times \cH_{\a' \ad'(2) \b'(2) \bd' \g \gd}(X_{12}) \, .
	\end{split}
\end{align} 
The tensor three-point function is encoded in the following polynomial:
\begin{align}
	\cH(X; U, V, W) = \cH_{ \a \ad(2) \b(2) \bd  \g \gd }(X) \, \boldsymbol{U}^{\a \ad(2)}  \boldsymbol{V}^{\b(2) \bd}  \boldsymbol{W}^{\g \gd} \, .
\end{align}
After solving \eqref{Diophantine equations}, we find the following linearly dependent polynomial structures:
%
\begin{align}
	\begin{split}
		&\big\{ Q_1 Q_2 Z_1 Z_2,P_3 Q_2 Z_2 \bar{P}_2,P_1 Z_1 Z_2 \bar{P}_2,P_3 Q_1 Q_2 \bar{P}_3,P_1 Q_1 Z_1 \bar{P}_3, \\
		& \hspace{10mm} P_1 P_3 \bar{P}_2
		\bar{P}_3, Q_1 Q_2 Q_3 \bar{Q}_3,Z_1 Z_2 Z_3 \bar{Q}_3,P_3 Q_2 \bar{P}_1 \bar{Q}_3,P_1 Z_1 \bar{P}_1 \bar{Q}_3, \\
		& \hspace{17mm} P_1 Q_3
		\bar{P}_2 \bar{Q}_3, P_2 Z_2 \bar{P}_2 \bar{Q}_3, P_2 Q_1 \bar{P}_3 \bar{Q}_3,P_3 Z_3 \bar{P}_3 \bar{Q}_3,Q_1 Z_1 \bar{Q}_1
		\bar{Q}_3, \\
		& \hspace{22mm} P_3 \bar{P}_2 \bar{Q}_1 \bar{Q}_3,Q_2 Z_2 \bar{Q}_2 \bar{Q}_3, P_1 \bar{P}_3 \bar{Q}_2 \bar{Q}_3,Q_3 Z_3
		\bar{Q}_3^2,P_2 \bar{P}_1 \bar{Q}_3^2,\bar{Q}_1 \bar{Q}_2 \bar{Q}_3^2 \big\}\,.
	\end{split}
\end{align}
Next we systematically apply the linear dependence relations \eqref{Linear dependence 1} to this list, which results in the following linearly independent structures:
%
\begin{align}
	\big\{P_3 Q_1 Q_2 \bar{P}_3,Q_1 Q_2 Q_3 \bar{Q}_3,P_3 Q_2 \bar{P}_1 \bar{Q}_3,P_1 Q_3 \bar{P}_2 \bar{Q}_3,P_2 Q_1 \bar{P}_3
	\bar{Q}_3,\bar{Q}_1 \bar{Q}_2 \bar{Q}_3^2\big\}\,.
\end{align}
We now construct the ansatz for this three-point function using the linearly independent structures above. After imposing conservation on all three points the final solution is
%
\begin{align} \label{QQbV}
	\begin{split}
		& \frac{A_1}{X^4}
		\Big(P_3 Q_1 Q_2 \bar{P}_3 + \sfrac{3}{2} P_1 Q_3 \bar{P}_2 \bar{Q}_3-\sfrac{3}{4} \bar{Q}_1 \bar{Q}_2\bar{Q}_3^2\Big) \\[1mm]
		& \hspace{5mm} +\frac{A_2}{X^4} \Big(P_1 Q_3 \bar{P}_2 \bar{Q}_3-\bar{Q}_1 \bar{Q}_2 \bar{Q}_3^2+Q_1 Q_2 Q_3 \bar{Q}_3\Big) \\[1mm]
		& \hspace{10mm} +\frac{A_3}{X^4} \Big(P_3 Q_2 \bar{P}_1 \bar{Q}_3-\sfrac{1}{2} P_1 Q_3 \bar{P}_2 \bar{Q}_3+P_2 Q_1 \bar{P}_3
		\bar{Q}_3+\sfrac{3}{4} \bar{Q}_1 \bar{Q}_2 \bar{Q}_3^2\Big)\,.
	\end{split}
\end{align}
Therefore we see that the correlation function $\langle J^{}_{(3/2,1)} \bar{J}'_{(3/2,1)} J''_{(1,0)} \rangle$ and, hence, $\langle Q \bar{Q} V \rangle$, is fixed up to three independent complex coefficients. After imposing the combined point-switch/reality condition on $Q$ and $\bar{Q}$, we find that the complex coefficients $A_{i}$ must be purely imaginary, i.e., $A_{i} = \text{i} \tilde{a}_{i}$. Hence, the correlation function $\langle Q \bar{Q} V \rangle$ is fixed up to three independent real parameters.

Next we determine the general structure of $\langle Q Q T \rangle$ and $\langle Q \bar{Q} T \rangle$, which are associated with the correlation functions $\langle J^{}_{(3/2,1)} J'_{(3/2,1)} J''_{(2,0)} \rangle$, $\langle J^{}_{(3/2,1)} \bar{J}'_{(3/2,1)} J''_{(2,0)} \rangle$ respectively using our general formalism. Since the number of structures grows rapidly with spin, we will simply present the final results after conservation. For $\langle J^{}_{(3/2,1)} J'_{(3/2,1)} J''_{(2,0)} \rangle$ we obtain a single independent structure (up to a complex coefficient):
%
\begin{align} \label{QQT}
	\begin{split}
		&\frac{A_1}{X^3} \Big(Q_1 Q_3 Q_2^2 \bar{P}_1+\sfrac{7}{4} P_3 Q_2^2 \bar{P}_1^2+\sfrac{1}{2} P_1 Q_3 Q_2 \bar{P}_1
		\bar{P}_2 \\
		& \hspace{10mm} -\sfrac{5}{4} Q_1 Q_3 Q_2 \bar{P}_2 \bar{Q}_1 -5 Q_1 Q_2 \bar{P}_3 \bar{Q}_1 \bar{Q}_2-\sfrac{7}{2} Q_2 \bar{P}_1
		\bar{Q}_1 \bar{Q}_2 \bar{Q}_3 \\
		& \hspace{20mm} +\sfrac{1}{2} P_1 Q_3 \bar{P}_2^2 \bar{Q}_1+\sfrac{7}{4} P_2 Q_1 \bar{P}_2 \bar{P}_3
		\bar{Q}_1-\sfrac{5}{4} \bar{P}_2 \bar{Q}_1^2 \bar{Q}_2 \bar{Q}_3 \Big)\,.
	\end{split}
\end{align}
This solution is manifestly compatible with the point-switch symmetry resulting from setting $J = J'$, hence, $\langle Q Q T \rangle$ is unique up to a complex parameter. On the other hand, for $\langle J^{}_{(3/2,1)} \bar{J}'_{(3/2,1)} J''_{(2,0)} \rangle$ we obtain four independent conserved structures proportional to complex coefficients
%
\begin{align} \label{QQbT}
	\begin{split}
		& \frac{A_1}{X^3} \Big( Q_1^2 Q_3 Q_2^2 + \sfrac{6}{7} P_1 Q_2 \bar{P}_1
		\bar{Q}_2 \bar{Q}_3 +\sfrac{6}{7} P_2 Q_1 \bar{P}_2 \bar{Q}_1 \bar{Q}_3 \\
		& \hspace{30mm} + \sfrac{6}{7} P_1 \bar{P}_2 \bar{Q}_1 \bar{Q}_2 \bar{Q}_3-\sfrac{10}{7} Q_1 Q_2 \bar{Q}_1 \bar{Q}_2 \bar{Q}_3\Big) \\[1mm]
		& +\frac{A_2}{X^3} \Big(P_2 Q_2 Q_1^2 \bar{P}_3+P_3 Q_2^2 Q_1 \bar{P}_1-P_2 Q_1 \bar{P}_2
		\bar{Q}_1 \bar{Q}_3 \\
		& \hspace{15mm} -P_1 Q_2 \bar{P}_1 \bar{Q}_2 \bar{Q}_3-P_1 \bar{P}_2 \bar{Q}_1 \bar{Q}_2 \bar{Q}_3+Q_2 Q_1 \bar{Q}_1 \bar{Q}_2 \bar{Q}_3 \Big) \\[1mm]
		&+\frac{A_3}{X^3} \Big(P_1 Q_1 Q_2 Q_3 \bar{P}_2-\sfrac{3}{7} P_2 Q_1 \bar{P}_2 \bar{Q}_1
		\bar{Q}_3-\sfrac{13}{14} P_1 \bar{P}_2 \bar{Q}_1 \bar{Q}_2 \bar{Q}_3 \\
		& \hspace{25mm} -\sfrac{3}{7} P_1 Q_2 \bar{P}_1 \bar{Q}_2
		\bar{Q}_3+\sfrac{3}{14} Q_1 Q_2 \bar{Q}_1 \bar{Q}_2 \bar{Q}_3 \Big) + \frac{A_4}{X^3} P_1^2 Q_3 \bar{P}_2^2 \,.
	\end{split}
\end{align}
After imposing the combined point-switch/reality condition, we find that the complex coefficients $A_{i}$ must be purely real. Hence, the three-point function $\langle Q \bar{Q} T \rangle$ is fixed up to four independent real parameters. The results \eqref{QQV}, \eqref{QQbV}, \eqref{QQT}, \eqref{QQbT} are in agreement with those found in \cite{Buchbinder:2022cqp}.



\subsection{General structure of three-point functions for arbitrary spins}\label{subsection3.2}


In four dimensions, three-point correlation functions of bosonic higher-spin conserved currents have been analysed in the following 
publications~\cite{Stanev:2012nq, Zhiboedov:2012bm} (see \cite{Osborn:1998qu,Kuzenko:1999pi, Buchbinder:2022kmj} for supersymmetric results). For three-point functions involving 
bosonic currents $J_{(s,0)} = J_{\a(s) \ad(s)}$, the general structure of the three-point function $\langle J^{}_{(s_{1},0)} J'_{(s_{2},0)} J''_{(s_{3},0)} \rangle$ was found to be fixed up to the 
following form \cite{Maldacena:2011jn, Giombi:2011rz, Zhiboedov:2012bm}:
\begin{equation}
	\langle  J^{}_{(s_{1},0)} J'_{(s_{2},0)} J''_{(s_{3},0)}  \rangle = \sum_{I=1}^{2 \min(s_{1}, s_{2}, s_{3}) + 1} a_{I} \, \langle  J^{}_{(s_{1},0)} J'_{(s_{2},0)} J''_{(s_{3},0)}  \rangle_{I} \, ,
\end{equation}
where $a_{I}$ are real coefficients and $\langle J^{}_{(s_{1},0)} J'_{(s_{2},0)} J''_{(s_{3},0)} \rangle_{I}$ are linearly independent conserved structures.\footnote{Note that if the reality condition is not imposed, the three-point function is fixed up to $2 \min(s_{i}) + 1$ structures with complex coefficients.} 
Among these $2 \min(s_{1}, s_{2}, s_{3}) + 1$ structures, $\min(s_{1}, s_{2}, s_{3}) + 1$ are parity-even while $\min(s_{1}, s_{2}, s_{3}) $ are parity odd. For correlation functions involving identical fields we must also impose point-switch symmetries. The following classification holds:
\begin{itemize}
	\item For three-point functions $\langle J^{}_{(s,0)} J'_{(s,0)} J''_{(s,0)} \rangle$ there are $2 s + 1$ conserved structures, $s+1$ being parity even and $s$ being parity odd.  
	When the fields coincide, i.e. $J = J' $ the number of structures is reduced to the $s+1$ parity-even structures in the case when the spin $s$ is even,
	or to the $s$ parity-odd structures in the case when $s$ is odd. 
		
	\item For three-point functions $\langle J^{}_{(s_{1},0)} J'_{(s_{1},0)} J''_{(s_{2},0)} \rangle$, there are $2 \min(s_{1}, s_{2}) + 1$ conserved structures, 
	 $\min(s_{1}, s_{2}) +1$ being parity even and $\min(s_{1}, s_{2}) $ being parity odd.  
	For $J = J'$, the number of structures is reduced to the $\min(s_{1}, s_{2})+1$ parity-even structures in the case when the spin $s_2$ is even, or to the $\min(s_{1}, s_{2})$ parity-odd structures in the case when $s_2$ is odd.
\end{itemize}
Note that the above classification is consistent with the results of \cite{Stanev:2012nq}, and we have explicitly reproduced them up to $s_{i} = 10$ in our computational approach. 

Now let us discuss three-point functions involving currents with $q = 1$, which define ``supersymmetry-like" fermionic higher-spin currents. 
The possible correlation functions that we can construct from these are 
$\langle J^{}_{(s_{1}, 1)} \, J'_{(s_{2}, 1)} \, J''_{(s_{3}, 0)}\rangle$ and $\langle J^{}_{(s_{1}, 1)} \, \bar{J}'_{(s_{2}, 1)} \, J''_{(s_{3}, 0)} \rangle$. 
Note that for $s_{1} = s_{2} = 3/2$ and $s_{3} = 1,2$ we obtain the familiar three-point functions \eqref{Susy current correlators - 1}. 
Based on our computational analysis we found that the three-point function $\langle J^{}_{(s_{1}, 1)} \, J'_{(s_{2}, 1)} \, J''_{(s_{3}, 0)}\rangle$ is fixed up 
to a unique structure after conservation in general. On the other hand, we found that three-point functions of the form $\langle J^{}_{(s_{1}, 1)} \, \bar{J}'_{(s_{2}, 1)} \, J''_{(s_{3}, 0)} \rangle$ 
are fixed up to $2 \min(s_{1}, s_{2}, s_{3}) + 1$ independent conserved structures. It's important to note that for these three-point functions there is no notion of parity-even/odd 
structures. 

We now dedicate the remainder of this section to classifying the number of independent structures in the general three-point functions
\begin{align} \label{Possible conserved three-point functions}
	\langle J^{}_{(s_{1}, q_{1})} \, J'_{(s_{2}, q_{2})} \, J''_{(s_{3}, q_{3})} \rangle \, , && \langle J^{}_{(s_{1}, q_{1})} \, \bar{J}'_{(s_{2}, q_{2})} \, J''_{(s_{3}, q_{3})} \rangle  \, ,
\end{align}
for arbitrary $(s_i, q_i)$. We investigated the general structure of these three-point functions up to $s_{i} = 10$. Provided that the inequalities \eqref{e1} are satisfied, we conjecture that the following classification holds in general:
\begin{itemize}
	\item For three-point functions $\langle J^{}_{(s_{1},q_{1})} J'_{(s_{2},q_{2})} J''_{(s_{3}, q_{3})} \rangle$, $\langle J^{}_{(s_{1}, q_{1})} \bar{J}'_{(s_{2}, q_{2})} J''_{(s_{3}, q_{3})} \rangle$ with $q_{1} \neq q_{2} \neq q_{3}$, there is a unique solution in general. Similarly, the three-point function is also unique for the cases: i) $q_{1} = 0$, $q_{2} \neq q_{3}$, and ii) $q_{1} = q_{2} = 0$ with $q_{3} \neq 0$.
	
	\item For three-point functions $\langle J^{}_{(s_{1},q)} J'_{(s_{2},q)} J''_{(s_{3},0)} \rangle$ there is a unique solution up to a complex coefficient. However, for the case where $s_{1} = s_{2}$ (fermionic or bosonic) and 
	$J = J'$, the structure survives the resulting point-switch symmetry only when $s_{3}$ is an 
	even integer.
	
	\item For three-point functions $\langle J^{}_{(s_{1},q)} \bar{J}'_{(s_{2},q)} J''_{(s_{3},0)} \rangle$ we obtain quite a non-trivial result which we will now explain.
	The number of structures, $N(s_{1}, s_{2}, s_{3};q)$, obeys the following formula:
	\begin{equation} \label{Number of structures}
		N(s_{1}, s_{2}, s_{3}; q) = 2 \min(s_{1}, s_{2}, s_{3}) + 1 - \text{max} \big( \tfrac{q}{2} - | s_{3} - \min(s_{1}, s_{2}) | , 0 \big) \, ,
	\end{equation}
	where $s_{1}, s_{2}$ are simultaneously integer/half-integer, for integer $s_{3}$.
	This formula can be arrived at using the following method. 
	Let us fix $s_{1}$, $s_{2}$ and let $q \geq 2$. By varying $s_{3}$ and computing the resulting conserved three-point function, 
	one can notice that if $s_3$ lies within the interval
	%
	\begin{equation} \label{weird interval}
	\min(s_{1}, s_{2}) - \frac{q}{2} < s_{3} < \min(s_{1}, s_{2}) + \frac{q}{2} \, ,
	\end{equation}
	then the number of structures is decreased from $2 \min(s_{1}, s_{2}, s_{3}) + 1$ by 
	\begin{equation}
	\d N(s_{1}, s_{2}, s_{3}; q)= \frac{q}{2} - | s_{3} - \min(s_{1}, s_{2}) |  \,. 
	\label{e2}
	\end{equation}
	For $s_{3}$ outside the interval \eqref{weird interval} there is always $2 \min(s_{1}, s_{2}, s_{3}) + 1$ structures in general.
	 It should also be noted that \eqref{Number of structures} is also valid for $q = 0,1$ (by virtue of the $\max()$ function). In these cases the additional 
	 term does not contribute and we obtain $N(s_{1}, s_{2}, s_{3}; 0) = N(s_{1}, s_{2}, s_{3}; 1) = 2 \min(s_{1}, s_{2}, s_{3}) + 1$.
	
	As examples, below we tabulate the number of structures in the conserved three-point functions 
	$\langle J^{}_{(s_{1},q)} \bar{J}'_{(s_{2},q)} J''_{(s_{3},0)} \rangle$ for some fixed $s_{1}, s_{2}$ while varying $q$ and $s_{3}$. 
	Let us recall that $q$ is necessarily even/odd when $s$ is integer/half-integer valued. In addition, since $J_{(s,q)} := J_{\a(s+\frac{q}{2}) \ad(s-\frac{q}{2})}$ it follows that the maximal allowed value of $q$ 
	in the above correlation function is  $2 \min(s_{1}, s_{2}) - 2$. Explicit solutions for particular cases are presented in appendix \ref{AppB}.
	
	\begin{center}
		\captionof{table}{No. of structures in $\langle J^{}_{(s_{1},q)} \bar{J}'_{(s_{2},q)} J''_{(s_{3},0)} \rangle$ for $s_{1} = 5$, $s_{2} = 6$.\label{Tab:T1}}
		\vspace{5mm}
		{\def\arraystretch{1.4}
		{\setlength{\tabcolsep}{3mm}
		\begin{tabular}{c*{10}{c}}
			\toprule
			& \multicolumn{9}{c}{$s_{3}$} \\
			\cmidrule(lr){2-10}
			\hspace{3mm} $q$ \hspace{3mm} & 1 & 2 & 3 & 4 & 5 & 6 & 7 & 8 & 9 \\
			\cmidrule(lr){1-1}
			\cmidrule(lr){2-10}
			0 & 3 & 5 & 7 & 9 & 11 & 11 & 11 & 11  & 11 \\[1mm]
			2 & 3 & 5 & 7 & 9 & \textcolor{red}{10} & 11 & 11 & 11  & 11 \\[1mm]
			4 & 3 & 5 & 7 & \textcolor{red}{8} & \textcolor{blue}{9} & \textcolor{red}{10} & 11 & 11 & 11\\[1mm]
			6 & 3 & 5 & \textcolor{red}{6} & \textcolor{blue}{7} & \textcolor{teal}{8} & \textcolor{blue}{9} & \textcolor{red}{10} & 11 & 11\\[1mm]
			8 & 3& \textcolor{red}{4} & \textcolor{blue}{5} & \textcolor{teal}{6} & \textcolor{violet}{7} & \textcolor{teal}{8} & \textcolor{blue}{9} & \textcolor{red}{10} & 11\\[1mm]
			\bottomrule
		\end{tabular}}}
	\caption*{ \textcolor{red}{$\bullet$} $\d N = 1$ \hspace{5mm} \textcolor{blue}{$\bullet$} $\d N = 2$ \hspace{5mm} \textcolor{teal}{$\bullet$} $\d N = 3$ \hspace{5mm} \textcolor{violet}{$\bullet$} $\d N = 4$}
	\end{center}

	\newpage
	
	\begin{center}
		\captionof{table}{No. of structures in $\langle J^{}_{(s_{1},q)} \bar{J}'_{(s_{2},q)} J''_{(s_{3},0)} \rangle$ for $s_{1} = 9/2$, $s_{2} = 11/2.$\label{Tab:T2}}
		\vspace{5mm}
		{\def\arraystretch{1.4}
		{\setlength{\tabcolsep}{3mm}
		\begin{tabular}{ c*{9}{c}}
			\toprule
			& \multicolumn{8}{c}{$s_{3}$} \\
			\cmidrule(lr){2-9}
			\hspace{3mm} $q$ \hspace{3mm} & 1 & 2 & 3 & 4 & 5 & 6 & 7 & 8 \\
			\cmidrule(lr){1-1}
			\cmidrule(lr){2-9}
			1 & 3 & 5 & 7 & 9 & 10 & 10 & 10 & 10 \\[1mm]
			3 & 3 & 5 & 7 & \textcolor{red}{8} & \textcolor{red}{9} & 10 & 10 & 10 \\[1mm]
			5 & 3 & 5 & \textcolor{red}{6} & \textcolor{blue}{7} & \textcolor{blue}{8} & \textcolor{red}{9} & 10 & 10 \\[1mm]
			7 & 3 & \textcolor{red}{4} & \textcolor{blue}{5} & \textcolor{teal}{6} & \textcolor{teal}{7} & \textcolor{blue}{8} & \textcolor{red}{9} & 10 \\[1mm]
			\bottomrule
		\end{tabular}}}
	\caption*{ \textcolor{red}{$\bullet$} $\d N = 1$ \hspace{5mm} \textcolor{blue}{$\bullet$} $\d N = 2$ \hspace{5mm} \textcolor{teal}{$\bullet$} $\d N = 3$}
	\end{center}

	The highlighted values are within the interval \eqref{weird interval} defined by $s_{1}$, $s_{2}$ and $q$, and we have used colour to identify the pattern in the number of structures. Analogous tables can be constructed for any choice of $s_{1}$, $s_{2}$ and it is easy to see that the results are consistent with the general formula \eqref{Number of structures}, which appears to hold for all such correlators within the bounds of our computational limitations $(s_{i} \leq 10)$.
	
	\item For three-point functions $\langle J^{}_{(s_{1},q)} \bar{J}'_{(s_{1},q)} J''_{(s_{2},0)} \rangle$ the number of structures adheres to the formula \eqref{Number of structures}. However, for $J = J'$, we must impose the combined point-switch/reality condition. After imposing this constraint we find that the free complex parameters must be purely real/imaginary for $s_{2}$ even/odd. 

\end{itemize}
The above classification appears to be complete, and we have not found any other permutations of fields/spins which give rise to new results.

\section*{Acknowledgements}
The authors are grateful to Sergei Kuzenko, Emmanouil Raptakis and Daniel Hutchings for valuable discussions. The work of E.I.B. is supported in part by the Australian Research Council, projects DP200101944
and DP230101629. The work of B.S. is supported by the \textit{Bruce and Betty Green Postgraduate Research Scholarship} under the Australian Government Research Training Program.



\appendix

\section{4D conventions and notation}\label{AppA}

Our conventions closely follow that of \cite{Buchbinder:1998qv}. For the Minkowski metric $\eta_{mn}$ we use the ``mostly plus'' convention: $\eta_{mn} = \text{diag}(-1,1,1,1)$. Spinor indices on spin-tensors are raised and lowered with the $\text{SL}(2,\mathbb{C})$ invariant spinor metrics
\begin{subequations}
	\begin{align}
		\ve_{\a \b} = 
		\begingroup
		\setlength\arraycolsep{4pt}
		\begin{pmatrix}
			\, 0 & -1 \, \\
			\, 1 & 0 \,
		\end{pmatrix}
		\endgroup 
		\, , & \hspace{10mm}
		\ve^{\a \b} =
		\begingroup
		\setlength\arraycolsep{4pt}
		\begin{pmatrix}
			\, 0 & 1 \, \\
			\, -1 & 0 \,
		\end{pmatrix}
		\endgroup 
		\, , \hspace{10mm}
		\ve_{\a \g} \, \ve^{\g \b} = \d_{\a}{}^{\b} \, , \\[4mm]
		\ve_{\ad \bd} = 
		\begingroup
		\setlength\arraycolsep{4pt}
		\begin{pmatrix}
			\, 0 & -1 \, \\
			\, 1 & 0 \,
		\end{pmatrix}
		\endgroup 
		\, , & \hspace{10mm}
		\ve^{\ad \bd} =
		\begingroup
		\setlength\arraycolsep{4pt}
		\begin{pmatrix}
			\, 0 & 1 \, \\
			\, -1 & 0 \,
		\end{pmatrix}
		\endgroup 
		\, , \hspace{10mm}
		\ve_{\ad \gd} \, \ve^{\gd \bd} = \d_{\ad}{}^{\bd} \, .
	\end{align}
\end{subequations}
Given the spinor fields $\f_{\a}$, $\bar{\f}_{\ad}$, the spinor indices $\a = 1, 2$, $\ad = \bar{1}, \bar{2}$ are raised and lowered according to the following rules:
\begin{align}
	\f_{\a} &= \ve_{\a \b} \, \f^{\b} \, , & \f^{\a} &= \ve^{\a \b} \, \f_{\b} \, , & \bar{\f}_{\ad} &= \ve_{\ad \bd} \, \bar{\f}^{\b} \, , & \bar{\f}^{\ad} &= \ve^{\ad \bd} \, \bar{\f}_{\bd} \, .
\end{align}
It is also useful to introduce the complex $2 \times 2$ $\s$-matrices, defined as follows:
\begin{align}
	\s_{0} &= 
	\begingroup
	\setlength\arraycolsep{4pt}
	\begin{pmatrix}
		\, 1 & 0 \, \\
		\, 0 & 1 \,
	\end{pmatrix}
	\endgroup 
	\, , & \hspace{5mm}
	\s_{1} &=
	\begingroup
	\setlength\arraycolsep{4pt}
	\begin{pmatrix}
		\, 0 & 1 \, \\
		\, 1 & 0 \,
	\end{pmatrix}
	\endgroup 
	\, , & \hspace{5mm}
	\s_{2} &=
	\begingroup
	\setlength\arraycolsep{4pt}
	\begin{pmatrix}
		\, 0 & -\text{i} \, \\
		\, \text{i} & 0 \,
	\end{pmatrix}
	\endgroup 
	\, , & \hspace{5mm}
	\s_{3} &=
	\begingroup
	\setlength\arraycolsep{4pt}
	\begin{pmatrix}
		\, 1 & 0 \, \\
		\, 0 & -1 \,
	\end{pmatrix}
	\endgroup 
	\, .
\end{align}
The $\s$-matrices span the Lie group $\text{SL}(2, \mathbb{C})$, the universal covering group of the Lorentz group $\text{SO}(3,1)$. Now let $\s_{m} = (\s_{0}, \vec{\s} )$, we denote the components of $\s_{m}$ as $(\s_{m})_{\a \ad}$, and define:
\begin{equation}
	(\tilde{\s}_{m})^{\ad \a} := \ve^{\ad \bd} \ve^{\a \b} (\s_{m})_{\b \bd} \, .
\end{equation}	
It can be shown that the $\s$-matrices possess the following useful properties:
\begin{subequations}
	\begin{align}
		(\s_{m} \tilde{\s}_{n} + \s_{n} \tilde{\s}_{m}  )_{\a}{}^{\b} &= - 2 \eta_{m n} \, \d_{\a}^{\b} \, , \\
		(\tilde{\s}_{m } \s_{n} + \tilde{\s}_{n} \s_{m}  )^{\ad}{}_{\bd} &= - 2 \eta_{m n} \, \d^{\ad}_{\bd} \, , \\
		\text{Tr}(\s_{m} \tilde{\s}_{n} ) &= - 2 \eta_{m n} \, , \\
		(\s^{m})_{\a \ad} (\tilde{\s}_{m})^{\bd \b} &= - 2 \d_{\a}^{\b} \d_{\ad}^{\bd} \, .
	\end{align}
\end{subequations}
The $\s$-matrices are then used to convert spacetime indices into spinor ones and vice versa according to the following rules:
\begin{equation}
	X_{\a \ad} = (\s^{m})_{\a \ad} X_{m} \, , \hspace{10mm} X_{m} = - \frac{1}{2} (\tilde{\s}_{m})^{\ad \a} X_{\a \ad} \, .
\end{equation}
\newpage

For imposing conservation equations on three-point functions, one must act on the generating function \eqref{Generating function} with the operators \eqref{Derivative operators}. For this, the following identities for the derivatives of the monomials $Q_{i}, Z_{i}$ are useful:
\begin{subequations}
	\begin{align}
		\pa_{X}^{\a \ad} Q_{1} &= - \frac{1}{X} \big( 2 v^{\a} \bar{w}^{\ad} + \hat{X}^{\ad \a} Q_{1} \big) \, ,\\
		\pa_{X}^{\a \ad} Q_{2} &= - \frac{1}{X} \big( 2 w^{\a} \bar{u}^{\ad} + \hat{X}^{\ad \a} Q_{2} \big) \, ,\\
		\pa_{X}^{\a \ad} Q_{3} &= - \frac{1}{X} \big( 2 u^{\a} \bar{v}^{\ad} + \hat{X}^{\ad \a} Q_{3} \big) \, ,
	\end{align}
\end{subequations}
\vspace{-8mm}
\begin{subequations}
	\begin{align}
		\pa_{X}^{\a \ad} Z_{1} &= - \frac{1}{X} \big( 2 u^{\a} \bar{u}^{\ad} + \hat{X}^{\ad \a} Z_{1} \big) \, ,\\
		\pa_{X}^{\a \ad} Z_{2} &= - \frac{1}{X} \big( 2 v^{\a} \bar{v}^{\ad} + \hat{X}^{\ad \a} Z_{2} \big) \, ,\\
		\pa_{X}^{\a \ad} Z_{3} &= - \frac{1}{X} \big( 2 w^{\a} \bar{w}^{\ad} + \hat{X}^{\ad \a} Z_{3} \big) \, .
	\end{align}
\end{subequations}
Analogous identities for derivatives of $\bar{Q}_{i}$ may be obtained by complex conjugation.


\section{Examples of three-point functions $\langle J^{}_{(s_{1},q)} \bar{J}'_{(s_{2},q)} J''_{(s_{3},0)} \rangle$}\label{AppB}

In this appendix we provide some examples of three-point functions $\langle J^{}_{(s_{1},q)} \bar{J}'_{(s_{2},q)} J''_{(s_{3},0)} \rangle$. In particular, we compute two of the examples presented in Tables \ref{Tab:T1} \& \ref{Tab:T2} to illustrate the decrease in the number of independent conserved structures for particular values of $q$. Due to the large size of the solutions for increasing $s_{i}$, we only present the simplest cases.

First let us consider the three-point function $\langle J^{}_{(s_{1},q)} \bar{J}'_{(s_{2},q)} J''_{(s_{3},0)} \rangle$ with $s_{1} = 5$, $s_{2} = 6$, $q = 8$ and $s_{3} = 2$. Using our formalism, all information about this correlation function is encoded in the following polynomial:
\begin{align}
	\cH(X; U, V, W) = \cH_{\a(1) \ad(9) \b(10) \bd(2) \g(2) \gd(2) }(X)  \, \boldsymbol{U}^{\a(1) \ad(9)}  \, \boldsymbol{V}^{\b(10) \bd(2)} \, \boldsymbol{W}^{\g(2) \gd(2)} \, .
\end{align}
There are 13 possible linearly independent structures that can be constructed in this case:
\begin{align}
	\begin{split}
		& \big\{ P_1 P_3 Q_1^2 Q_2 \bar{P}_3^2 \bar{Q}_3^6,P_1 Q_1^2 Q_2 Q_3 \bar{P}_3 \bar{Q}_3^7, P_3 Q_1 Q_2 \bar{P}_1 \bar{Q}_1 \bar{Q}_3^8, \\
		& \hspace{5mm} P_3 Q_1^2 Q_2 \bar{P}_3 \bar{Q}_1 \bar{Q}_3^7, P_1 Q_1 Q_2 Q_3 \bar{P}_1 \bar{Q}_3^8, P_1 P_3 Q_1 Q_2 \bar{P}_1 \bar{P}_3 \bar{Q}_3^7, \\
		& \hspace{10mm} P_1 Q_1 Q_3 \bar{P}_2 \bar{Q}_1 \bar{Q}_3^8,P_2 Q_1^2
		\bar{P}_3 \bar{Q}_1 \bar{Q}_3^8,P_1 \bar{P}_1 \bar{Q}_1 \bar{Q}_2 \bar{Q}_3^9, Q_1 \bar{Q}_1^2 \bar{Q}_2 \bar{Q}_3^9, \\
		& \hspace{15mm} P_1
		P_3 Q_2 \bar{P}_1^2 \bar{Q}_3^8, Q_1^2 Q_2 Q_3 \bar{Q}_1
		\bar{Q}_3^8, P_1^2 Q_3 \bar{P}_1 \bar{P}_2 \bar{Q}_3^8 \big\} \, .
	\end{split}
\end{align}
We now impose conservation on all three points. The following solution is obtained:
\begin{align}
	\begin{split}
		& \frac{A_1}{X^{11}} \Big(\frac{22}{189} Q_1 \bar{Q}_1^2 \bar{Q}_2 \bar{Q}_3^9-\frac{11}{63} P_1 \bar{P}_1 \bar{Q}_1 \bar{Q}_2
		\bar{Q}_3^9+\frac{11}{56} P_1^2 Q_3 \bar{P}_1 \bar{P}_2 \bar{Q}_3^8 \\
		& \hspace{15mm} -\frac{121}{378} Q_1^2 Q_2 Q_3 \bar{Q}_1
		\bar{Q}_3^8-\frac{143}{756} P_1 Q_1 Q_3 \bar{P}_2 \bar{Q}_1 \bar{Q}_3^8-\frac{11}{14} P_1 Q_1^2 Q_2 Q_3
		\bar{P}_3 \bar{Q}_3^7 \\
		& \hspace{25mm}+\frac{1}{2} P_3 Q_1^2 Q_2 \bar{P}_3 \bar{Q}_1 \bar{Q}_3^7+P_1 P_3 Q_1^2 Q_2 \bar{P}_3^2
		\bar{Q}_3^6 \Big) \\[2mm]
		& + \frac{A_2}{X^{11}} \Big(\frac{31}{54} Q_1 \bar{Q}_1^2 \bar{Q}_2 \bar{Q}_3^9-\frac{11}{18} P_1 \bar{P}_1
		\bar{Q}_1 \bar{Q}_2 \bar{Q}_3^9+\frac{11}{16} P_1^2 Q_3 \bar{P}_1 \bar{P}_2 \bar{Q}_3^8 \\
		& \hspace{15mm} -\frac{10}{27} Q_1^2 Q_2 Q_3 \bar{Q}_1 \bar{Q}_3^8-\frac{143}{216} P_1 Q_1 Q_3 \bar{P}_2 \bar{Q}_1 \bar{Q}_3^8-\frac{1}{4} P_2 Q_1^2 \bar{P}_3 \bar{Q}_1 \bar{Q}_3^8 \\
		& \hspace{25mm} -\frac{1}{4} P_1 Q_1^2 Q_2 Q_3 \bar{P}_3 \bar{Q}_3^7+P_1 P_3 Q_1 Q_2 \bar{P}_1
		\bar{P}_3 \bar{Q}_3^7-\frac{3}{4} P_3 Q_1^2 Q_2 \bar{P}_3 \bar{Q}_1 \bar{Q}_3^7 \Big) \\[2mm]
		& + \frac{A_3}{X^{11}}  \Big( \frac{2}{3} Q_1
		\bar{Q}_1^2 \bar{Q}_2 \bar{Q}_3^9-P_1 \bar{P}_1 \bar{Q}_1 \bar{Q}_2 \bar{Q}_3^9+P_1 Q_1 Q_2 Q_3 \bar{P}_1
		\bar{Q}_3^8 \\
		& \hspace{25mm} + P_1^2 Q_3 \bar{P}_1 \bar{P}_2 \bar{Q}_3^8-\frac{2}{3} Q_1^2 Q_2 Q_3 \bar{Q}_1
		\bar{Q}_3^8-\frac{2}{3} P_1 Q_1 Q_3 \bar{P}_2 \bar{Q}_1 \bar{Q}_3^8\Big) \\[2mm]
		& + \frac{A_4}{X^{11}} \Big( \frac{43}{27} Q_1
		\bar{Q}_1^2 \bar{Q}_2 \bar{Q}_3^9-\frac{8}{9} P_1 \bar{P}_1 \bar{Q}_1 \bar{Q}_2 \bar{Q}_3^9+P_1 P_3 Q_2
		\bar{P}_1^2 \bar{Q}_3^8 \\
		& \hspace{15mm} +P_1^2 Q_3 \bar{P}_1 \bar{P}_2 \bar{Q}_3^8-\frac{97}{54} Q_1^2 Q_2 Q_3 \bar{Q}_1
		\bar{Q}_3^8-2 P_3 Q_1 Q_2 \bar{P}_1 \bar{Q}_1 \bar{Q}_3^8 \\
		& \hspace{25mm} -\frac{44}{27} P_1 Q_1 Q_3 \bar{P}_2 \bar{Q}_1
		\bar{Q}_3^8-\frac{1}{2} P_2 Q_1^2 \bar{P}_3 \bar{Q}_1 \bar{Q}_3^8 \Big) \, ,
	\end{split}
\end{align}
where $A_{i}$ are complex coefficients. Hence we see that this three-point function is fixed up to four independent conserved structures. Recall that for $q = 0$, the three-point function $\langle J^{}_{(s_{1},q)} \bar{J}'_{(s_{2},q)} J''_{(s_{3},0)} \rangle$ reduces to a three-point function of vector-like currents. Hence, we should expect $2 \min(s_{1}, s_{2}, s_{3})+ 1 = 5$ independent structures. Similar results can be obtained for other values of $q$ and $s_{3}$, which are contained in Table \ref{Tab:T1}.

Next, let us consider the three-point function $\langle J^{}_{(s_{1},q)} \bar{J}'_{(s_{2},q)} J''_{(s_{3},0)} \rangle$ with $s_{1} = 9/2$, $s_{2} = 11/2$, $q = 7$ and $s_{3} = 2$. All information about this correlation function is encoded in the following polynomial:
\begin{align}
	\cH(X; U, V, W) = \cH_{\a(1) \ad(8) \b(9) \bd(2) \g(2) \gd(2) }(X)  \, \boldsymbol{U}^{\a(1) \ad(8)}  \, \boldsymbol{V}^{\b(9) \bd(2)} \, \boldsymbol{W}^{\g(2) \gd(2)} \, .
\end{align}
In this case there are also 13 possible linearly independent structures:
\begin{align}
	\begin{split}
		& \big\{ P_1 P_3 Q_1^2 Q_2 \bar{P}_3^2 \bar{Q}_3^5,P_1 Q_1^2 Q_2 Q_3 \bar{P}_3 \bar{Q}_3^6, P_1
		P_3 Q_2 \bar{P}_1^2 \bar{Q}_3^7, \\
		& \hspace{8mm} P_3 Q_1^2 Q_2 \bar{P}_3 \bar{Q}_1 \bar{Q}_3^6,P_1 Q_1 Q_2 Q_3 \bar{P}_1 \bar{Q}_3^7, P_1 Q_1 Q_3 \bar{P}_2 \bar{Q}_1 \bar{Q}_3^7,   \\
		& \hspace{15mm} Q_1^2 Q_2 Q_3 \bar{Q}_1 \bar{Q}_3^7,P_3 Q_1 Q_2 \bar{P}_1 \bar{Q}_1 \bar{Q}_3^7, P_1 P_3 Q_1 Q_2 \bar{P}_1 \bar{P}_3 \bar{Q}_3^6, \\
		& \hspace{20mm} P_2 Q_1^2 \bar{P}_3 \bar{Q}_1 \bar{Q}_3^7,P_1 \bar{P}_1 \bar{Q}_1 \bar{Q}_2 \bar{Q}_3^8, P_1^2 Q_3 \bar{P}_1 \bar{P}_2 \bar{Q}_3^7, Q_1 \bar{Q}_1^2 \bar{Q}_2 \bar{Q}_3^8 \big\} \, .
	\end{split}
\end{align}
We now impose conservation on all three points, and the following solution is obtained:
\begin{align}
	\begin{split}
		& \frac{A_{1}}{X^{10}} \Big( \frac{5}{36} Q_1 \bar{Q}_1^2 \bar{Q}_2 \bar{Q}_3^8-\frac{5}{24} P_1 \bar{P}_1 \bar{Q}_1 \bar{Q}_2
		\bar{Q}_3^8+\frac{5}{21} P_1^2 Q_3 \bar{P}_1 \bar{P}_2 \bar{Q}_3^7 \\
		& \hspace{15mm}-\frac{25}{72} Q_1^2 Q_2 Q_3 \bar{Q}_1
		\bar{Q}_3^7-\frac{115}{504} P_1 Q_1 Q_3 \bar{P}_2 \bar{Q}_1 \bar{Q}_3^7-\frac{5}{6} P_1 Q_1^2 Q_2 Q_3 \bar{P}_3
		\bar{Q}_3^6 \\
		& \hspace{25mm} + \frac{1}{2} P_3 Q_1^2 Q_2 \bar{P}_3 \bar{Q}_1 \bar{Q}_3^6+P_1 P_3 Q_1^2 Q_2 \bar{P}_3^2
		\bar{Q}_3^5 \Big) \\[2mm]
		& +\frac{A_{2}}{X^{10}} \Big(\frac{7}{12} Q_1 \bar{Q}_1^2 \bar{Q}_2 \bar{Q}_3^8-\frac{5}{8} P_1 \bar{P}_1
		\bar{Q}_1 \bar{Q}_2 \bar{Q}_3^8+\frac{5}{7} P_1^2 Q_3 \bar{P}_1 \bar{P}_2 \bar{Q}_3^7 \\
		& \hspace{15mm} -\frac{3}{8} Q_1^2 Q_2 Q_3 \bar{Q}_1 \bar{Q}_3^7-\frac{115}{168} P_1 Q_1 Q_3 \bar{P}_2 \bar{Q}_1 \bar{Q}_3^7-\frac{1}{4} P_2 Q_1^2
		\bar{P}_3 \bar{Q}_1 \bar{Q}_3^7 \\
		& \hspace{25mm} -\frac{1}{4} P_1 Q_1^2 Q_2 Q_3 \bar{P}_3 \bar{Q}_3^6+P_1 P_3 Q_1 Q_2 \bar{P}_1
		\bar{P}_3 \bar{Q}_3^6-\frac{3}{4} P_3 Q_1^2 Q_2 \bar{P}_3 \bar{Q}_1 \bar{Q}_3^6\Big) \\[2mm]
		& +\frac{A_{3}}{X^{10}} \Big(\frac{2}{3} Q_1
		\bar{Q}_1^2 \bar{Q}_2 \bar{Q}_3^8-P_1 \bar{P}_1 \bar{Q}_1 \bar{Q}_2 \bar{Q}_3^8+P_1 Q_1 Q_2 Q_3 \bar{P}_1
		\bar{Q}_3^7 \\
		& \hspace{15mm} +P_1^2 Q_3 \bar{P}_1 \bar{P}_2 \bar{Q}_3^7-\frac{2}{3} Q_1^2 Q_2 Q_3 \bar{Q}_1
		\bar{Q}_3^7-\frac{2}{3} P_1 Q_1 Q_3 \bar{P}_2 \bar{Q}_1 \bar{Q}_3^7\Big) \\[2mm]
		& +\frac{A_{4}}{X^{10}} \Big(\frac{19}{12} Q_1
		\bar{Q}_1^2 \bar{Q}_2 \bar{Q}_3^8-\frac{7}{8} P_1 \bar{P}_1 \bar{Q}_1 \bar{Q}_2 \bar{Q}_3^8+P_1 P_3 Q_2
		\bar{P}_1^2 \bar{Q}_3^7 \\
		& \hspace{15mm} +P_1^2 Q_3 \bar{P}_1 \bar{P}_2 \bar{Q}_3^7-\frac{43}{24} Q_1^2 Q_2 Q_3 \bar{Q}_1
		\bar{Q}_3^7-2 P_3 Q_1 Q_2 \bar{P}_1 \bar{Q}_1 \bar{Q}_3^7 \\
		& \hspace{25mm} -\frac{13}{8} P_1 Q_1 Q_3 \bar{P}_2 \bar{Q}_1 \bar{Q}_3^7-\frac{1}{2} P_2 Q_1^2 \bar{P}_3 \bar{Q}_1 \bar{Q}_3^7\Big) \, ,
	\end{split}
\end{align}
where $A_{i}$ are complex coefficients. Hence we see that this three-point function is fixed up to four independent conserved structures. Recall that for the $q = 1$ case we expect $2 \min(s_{1}, s_{2}, s_{3})+ 1 = 5$ independent structures. Similar results are obtained for other values of $q$ and $s_{3}$, and with further testing we obtain Table \ref{Tab:T2}.

\section{Three-point functions involving scalars and spinors}\label{AppC}

In this appendix we provide some examples of three-point functions involving scalars, spinors and a conserved tensor operator. The results here serve as a consistency check against those presented in \cite{Osborn:1993cr,Elkhidir:2014woa}.

\newpage
\noindent
\textbf{Correlation function} $\langle O \, O' J_{(s,0)} \rangle$\textbf{:}\\[2mm]
Let $O$, $O'$ be scalar operators of dimension $\D_{1}$ and $\D_{2}$ respectively. We consider the three-point function $\langle O \, O' J_{(s,0)} \rangle$. According to the formula \eqref{e1}, a three-point function can be constructed only if $J$ is in the $(s,s)$ representation. Using the general formula, the ansatz for this three-point function is:
\begin{align}
		\langle O(x_{1}) \, O'(x_{2}) \, J_{\g(s) \gd(s)}(x_{3}) \rangle &= \frac{ 1 }{(x_{13}^{2})^{\D_{1}} (x_{23}^{2})^{\D_{2}}} \, \cH_{\g(s) \gd(s)}(X_{12}) \, .
\end{align} 
All information about this correlation function is encoded in the following polynomial:
\begin{align}
	\cH(X; W) = \cH_{ \g(s) \gd(s) }(X)  \, \boldsymbol{W}^{\g(s) \gd(s)} \, .
\end{align}
We recall that $\cH$ satisfies the homogeneity property $\cH(X) = X^{s+2 - \D_{1} - \D_{2}} \hat{\cH}(X)$, where $\hat{\cH}(X)$ is homogeneous degree $0$. The only possible structure for $\hat{\cH}(X)$ is:
\begin{align}
	\hat{\cH}(X; W) = A \, Z_{3}^{s} \, .
\end{align}
where $A$ is a complex coefficient. After imposing conservation on $x_{3}$ using the methods outlined in subsection \ref{subsubsection2.2.2}, we find 
\begin{align}
	D_{3} \tilde{\cH}(X; W) = A \left(\D_1-\D _2\right) (-1)^{s+1} s (s+1) Z_3^{s-1} X^{\D _1-\D_2-s-3} \, .
\end{align}
Hence, we find that this three-point function is compatible with conservation on $x_{3}$ only for $\D_{1} = \D_{2}$. When the scalars $O$, $O'$ coincide, then the solution satisfies the point-switch symmetry associated with exchanging $x_{1}$ and $x_{2}$ only for even $s$. This result is in agreement with \cite{Elkhidir:2014woa}. \\[5mm]
\noindent
\textbf{Correlation function} $\langle \psi \, \bar{\psi}' J_{(s,q)} \rangle$\textbf{:}\\[2mm]
Let $\psi$, $\bar{\psi}'$ be spinor operators of dimension $\D_{1}$ and $\D_{2}$ respectively. We now consider the three-point function $\langle \psi \, \bar{\psi}' J_{(s,q)} \rangle$. According to the formula \eqref{e1}, a three-point function can be constructed only if $J$ belongs to the representations $(s,s)$, $(s-1,s+1)$ or $(s+1,s-1)$ (the latter two corresponding to $q = 2$). First consider the $(s,s)$ representation. Using the general formula, the ansatz for this three-point function is:
\begin{align}
	\langle \psi_{\a}(x_{1}) \, \bar{\psi}_{\bd}'(x_{2}) \, J_{\g(s) \gd(s)}(x_{3}) \rangle &= \frac{ \cI_{\a}{}^{\ad'}(x_{13}) \, \bar{\cI}_{\bd}{}^{\b'}(x_{23}) }{(x_{13}^{2})^{\D_{1}} (x_{23}^{2})^{\D_{2}}} \, \cH_{\ad' \b' \g(s) \gd(s)}(X_{12}) \, .
\end{align} 
All information about this correlation function is encoded in the following polynomial:
\begin{align}
	\cH(X; U, V, W) = \cH_{ \ad \b \g(s) \gd(s) }(X)  \, \boldsymbol{U}^{\ad}  \boldsymbol{V}^{\b } \boldsymbol{W}^{\g(s) \gd(s)} \, .
\end{align}
We recall that $\cH$ satisfies the homogeneity property $\cH(X) = X^{s+2 - \D_{1} - \D_{2}} \hat{\cH}(X)$, where $\hat{\cH}(X)$ is homogeneous degree $0$. In this case there are two possible linearly independent structures for $\hat{\cH}(X)$:
\begin{align}
	\hat{\cH}(X; U, V, W) = A_{1} P_{2} \bar{P}_{1} Z_{3}^{s-1}  + A_{2} \, Q_{1} \bar{Q}_{2} Z_{3}^{s-1} \, ,
\end{align}
where $A_{1}$ and $A_{2}$ are complex coefficients. After imposing conservation on $x_{3}$ using the methods outlined in subsection \ref{subsubsection2.2.2}, we find 
\begin{align}
	\begin{split}
		D_{3} \tilde{\cH}(X; U,V,W) &= \left(\D _1-\D_2\right) (-1)^{s+1}  \big\{
		(A_1 + (s^2+s-1 ) A_2 ) \, Q_2 \bar{P}_1 \\
		& \hspace{20mm} + ( (s^2+s-1) A_1 + A_2 ) \bar{P}_2 \bar{Q}_1 \big\} Z_3^{s-2} X^{\D _1-\D_2-s-3} \, .
	\end{split}
\end{align}
Hence, we find that this three-point function is automatically compatible with conservation on $x_{3}$ for $\D_{1} = \D_{2}$. For $\D_{1} \neq \D_{2}$ it is simple to see that conservation is satisfied only for $s=1$, which results in $A_{1} = - A_{2}$ and, hence, the solution is unique. However, for $s>1$ there is no solution in general. In the case where $\psi = \psi'$, we also have to impose the combined point-switch/reality condition, which results in the coefficients $A_{i}$ being purely real/imaginary for $s$ even/odd. This result is consistent with \cite{Elkhidir:2014woa}.

Now let us consider the $(s+1,s-1)$ representation, with $s > 1$. Note that the analysis for $(s-1, s+1)$ is essentially identical and will be omitted. Using the general formula, the ansatz for this three-point function is:
\begin{align}
	\langle \psi_{\a}(x_{1}) \, \bar{\psi}_{\bd}'(x_{2}) \, J_{\g(s+1) \gd(s-1)}(x_{3}) \rangle &= \frac{ \cI_{\a}{}^{\ad'}(x_{13}) \, \bar{\cI}_{\bd}{}^{\b'}(x_{23}) }{(x_{13}^{2})^{\D_{1}} (x_{23}^{2})^{\D_{2}}} \, \cH_{\ad' \b' \g(s+1) \gd(s-1)}(X_{12}) \, .
\end{align} 
All information about this correlation function is encoded in the following polynomial:
\begin{align}
	\cH(X; U, V, W) = \cH_{ \ad \b \g(s+1) \gd(s-1) }(X)  \, \boldsymbol{U}^{\ad}  \boldsymbol{V}^{\b } \boldsymbol{W}^{\g(s+1) \gd(s-1)} \, .
\end{align}
We recall that $\cH$ satisfies the homogeneity property $\cH(X) = X^{s+2 - \D_{1} - \D_{2}} \hat{\cH}(X)$, where $\hat{\cH}(X)$ is homogeneous degree $0$. In this case there is only one possible structure for $\hat{\cH}(X)$:
\begin{align}
	\hat{\cH}(X; U, V, W) = A \, P_{2} \bar{Q}_{1} Z_{3}^{s-1}  \, ,
\end{align}
where $A$ is a complex coefficient. After imposing conservation on $x_{3}$ using the methods outlined in subsection \ref{subsubsection2.2.2}, we find 
\begin{align}
		D_{3} \tilde{\cH}(X;  U, V, W) &= A \left(\D_1-\D _2-1 \right) (-1)^{s+1} ( s^2 + s - 2 ) \, \bar{P}_1 \bar{P}_2 Z_3^{s-2}
		X^{\D _1-\D _2-s-3} \, .
\end{align}
Hence, we find that this three-point function is automatically compatible with conservation on $x_{3}$ for $\D_{2} = \D_{1} - 1$. For $\D_{2} \neq \D_{1} - 1$ there is no solution in general (recall that $s>1$). This result is also consistent with \cite{Elkhidir:2014woa}.



\printbibliography[heading=bibintoc,title={References}]



\end{document}